%% file: 00main.tex
\documentclass[acmsmall]{acmart}

\AtBeginDocument{%
  \providecommand\BibTeX{{%
    \normalfont B\kern-0.5em{\scshape i\kern-0.25em b}\kern-0.8em\TeX}}}

\setcopyright{rightsretained}
\acmJournal{PACMHCI}
\acmYear{2022} \acmVolume{6} \acmNumber{CSCW2} \acmArticle{552} \acmMonth{11} \acmPrice{}\acmDOI{10.1145/3555610}

\copyrightyear{2021}

\acmISBN{978-1-4503-XXXX-X/18/06}

\usepackage{xcolor} 
\usepackage{soul}
\usepackage{multirow}

\newcommand{\textadd}[1]{\textcolor{black}{#1}} % purple
\begin{document}


% \title{Challenges in Conducting HCI Research in Collaborative Academic HCI and Health Research Teams in the US: a Team Science Perspective}
\title{Using HCI in Cross-Disciplinary Teams: A Case Study of Academic Collaboration in HCI-Health Teams in the US Using a Team Science Perspective }

\author{Elena Agapie}
\email{eagapie@uci.edu}
\affiliation{%
  \institution{University of California, Irvine}
  \state{California}
  \city{Irvine}
  \country{USA}}

\author{Shefali Haldar}
\email{shefali.haldar@merck.com}
\affiliation{%
  \institution{Merck \& Co., Inc.}
  \state{Massachusetts}
  \city{Boston}
  \country{USA}}

\author{Sharmaine Galvez Poblete}
\affiliation{%
  \institution{University of California, Irvine}
  \state{California}
  \city{Irvine}
  \country{USA}}
\email{spoblete@uci.edu}

\renewcommand{\shortauthors}{Agapie, Haldar, Poblete}

\begin{abstract}
  \input{00abstract.tex}

\end{abstract}



\maketitle
\input{01introduction.tex}

\input{02relatedwork.tex}

\input{03methods.tex}
\input{04results.tex}

\input{05discussion.tex}

\input{06limitations-conclusion.tex}
% \input{07-Elena-editing.tex}

% \input{sample-removedtext-EA.tex}

\bibliographystyle{ACM-Reference-Format}
\bibliography{10references}

\appendix

\end{document}

%% file: 00abstract.tex
Human-centered computing research has been increasingly applied to address important challenges in the health domain. Conducting research in cross-disciplinary teams can come with a lot of challenges in integrating knowledge across fields. Yet, we do not know what challenges HCI researchers encounter in building collaborations with health researchers, and how these researchers negotiate challenges while balancing their professional goals. We interviewed 17 early- and mid-career HCI faculty working in the United States who conducted research in collaboration with health researchers. Drawing from a Team Science framework, we share participants' lived experiences and identify major challenges that HCI researchers encounter when finding, collaborating with, and negotiating with health collaborators when building technologies. We propose ways to better support research collaboration aimed at designing technologies using human-centered computing approaches. This includes strategies to support HCI researchers at individual, institutional, research community, and funding agencies levels through tools to translate disciplinary approaches. We suggest institutional policies to support HCI researchers through training, networking, and promotion.

%% file: 01introduction.tex
\section{Introduction}

The current scientific environment promotes collaboration across scientific fields \cite{national2015enhancing}. Team-based scientific research can increase research impact, novelty, reach, and productivity \cite{national2015enhancing}. However, cross-disciplinary teams can experience barriers in achieving research that integrates multiple perspectives such as having different goals, research approaches, publication practices, funding approaches, organizational support, academic and career recognition, or promotion practices that recognize cross-disciplinary work \cite{national2015enhancing}. 

CSCW research has historically studied some of the challenges and collaborative practices of scientific teams towards the production of scientific knowledge with the special focus and goal of technology design. For example, Olson et al. studied remote collaboration \cite{olson2000distance,olson2013working} and knowledge integration \cite{olson2000distance,olson2013working}; Kraut et al. identified the role of relationships and tasks in collaboration \cite{kraut1987relationships}; and Lee et al. showed how work practices \cite{rolland2014work}, human and cyberinfrastructure, support scientific collaboration \cite{bietz2012adapting,bietz2012sustaining,lee2006human}. 

Although past CSCW work has carefully studied team practices, there is limited research on the holistic, sociotechnical challenges that research teams in HCI encounter when conducting cross-disciplinary research aimed at creating technological artifacts. 
% Early career researchers face additional hurdles in (example, example). 
We know little about how these challenges affect the individuals and impact their teams, how institutions influence these teams' success, and how researchers must adjust their full research life-cycle -- from research conception to research dissemination and translation -- in response to these challenges.   

One such instance of cross-disciplinary collaboration is the rapid uptake of human-centered computing approaches in the health domain. This intersection of HCI and Health research strives to address important health challenges through technology, such as engaging patients in their care through peer-support communities \cite{lazar2019parting, huh2014weaving,haldar2020patient,o2018suddenly,nakikj2017park}, patient-generated health apps \cite{schroeder2018examining,chung2016boundary,saksono2021storymap,doherty2019engagement}, and health information portals	\cite{hong2017supporting,latulipe2015design,cajander2019electronic}. Researchers have aimed to support the documentation, coordination, and decision-making work of healthcare providers \cite{kulp2019comparing,cai2019hello,jacobs2021designing,moody2004electronic,rothschild2005leveraging}, and health services such as patient-centric services, safety, and outcomes \cite{joseph2006care,kobb2003enhancing,harper2005review,edwards2004reducing}. Cross-disciplinary teams of HCI and health researchers have also explored technologies to address challenges people have in managing chronic conditions \cite{mishra2019supporting,macleod2017grateful,kuo2019development,seo2019balancing}, acute events \cite{pollack2016closing,zhang2021mapping,verdezoto2020indigenous}, or health within everyday and clinical contexts \cite{kaziunas2019precarious,jacobs2018mypath,beede2020human,karkar2017tummytrials}. 
% Further, \textchange{human centered design} research is increasingly used to develop health interventions and make treatments more efficacious \cite{mohr2013behavioral, lyon2016user}. 
Health communities have adopted human-centered approaches or are increasingly doing so in fields such as the Medical Informatics, Implementation Science \cite{lyon2016user,lyon2020leveraging}, or Mental Health \cite{kruzan2021centering,mohr2013behavioral}. Cross-disciplinary research between human-centered computing and health has increasingly been supported by US public funding agencies \cite{nih-hci,nsf-sch}, leading to almost 500 projects that use human-centered design approaches funded to date by the National Institutes of Health (NIH) \cite{nih-hci}, and several research and training centers introducing human-centered methods to mental health research \cite{nih-training-center}.

To explore the collaborative experiences of HCI researchers working in cross-disciplinary teams, we focus on a case study of collaborations within teams of HCI-Health researchers in principal investigator roles and the factors that impact their team's collaboration. Within HCI and Health, researchers have been documenting personal experiences of challenges in collaboration, comparing methods and practices of HCI research with those of Health research \cite{blandford2015strategies,blandford2018lessons,blandford2018seven}, surfacing differences in methods \cite{singh2017hci,blandford2018seven,buis2020common,blandford2019hci4health} and research practices \cite{singh2017hci,blandford2018seven,buis2020common,furniss2014fieldwork,blandford2019hci4health}. However, there is limited understanding of the challenges that HCI-Health teams encounter in conducting cross-disciplinary research. 

To understand the HCI-Health collaboration challenges, we focus on HCI investigators who are early to mid-career because of the unique experiences they face in establishing cross-disciplinary teams, gaining recognition in multiple fields with limited training as a primary investigators, and planning their long-term career trajectories around this type of research \cite{national2015enhancing,spring2012emerging,cukier2020team}. Such difficulties can cause divisiveness in a team and even lead researchers to give up on cross-disciplinary research \cite{salazar2012facilitating, campbell2005overcoming}. Ultimately, a lack of early-career training and support can harm the success of cross-disciplinary research, limiting the innovation that might result from the collaboration \cite{national2015enhancing}. Understanding the experiences that these researchers face in HCI-Health teams is key to identifying solutions that can address these challenges.

We interviewed HCI researchers working at a variety of universities (Research R1 and R2 \cite{shulman2001carnegie} and Liberal Arts Colleges) in the United States about their experiences working in HCI-Health cross-disciplinary teams, what challenges they encountered, and how they navigated them. We recruited 17 early-career faculty (non-tenured) and faculty who recently transitioned out of early career (recently tenured) from 16 universities. 

Because organizational structures, resources, and promotion policies are all factors that impact cross-disciplinary research collaboration \cite{national2015enhancing}, we scope our research to US universities, a complex environment with variability between different types of institutions (e.g. resources of R1 universities vs. non-research universities, different promotion policies), and some fixed aspects of the infrastructure (e.g. types of public funding, the existence of tenure).

% Achieving the goals of cross-disciplinary research and innovating across disciplines relies on \textit{transdisciplinary research}, or tight integration of perspectives across different fields. Integrating perspectives across scientific disciplines is difficult.

We draw on Team Science to study collaborations, a field that draws from CSCW research, as well as psychology, organizational sciences, management, science and technology studies, and more \cite{national2015enhancing,hall2018science}. Team Science takes a holistic approach to investigate the factors that impact collaboration at multiple levels of scale, and proposed calls to action for the different stakeholders that impact collaboration: individual, team, organization, research community, and funding agencies \cite{national2015enhancing}. Team science can offer the CSCW community a holistic perspective to investigating how human-centered technology design practices are incorporated in cross-disciplinary collaborations, how collaboration challenges manifest, and how we can draw on existing guidelines to help support collaborations involving the use of human-centered computing approaches in other fields. Specifically, we investigate challenges in integrating human-centered approaches in HCI-Health teams that are designing technology.

Our research also contributes to the growing body of research that examines practices in the HCI community and proposes calls to action for how to improve community practices to reduce disparities \cite{ogbonnaya2020critical}, be inclusive of stakeholder perspectives that are typically excluded \cite{hofmann2020living,liang2021embracing,ogbonnaya2020critical}, improve ethical practices \cite{vitak2016beyond}, and broaden dissemination of research \cite{smith2018breaking,smith2020disseminating}. We contribute how a focus on research integration, through translational resources, institutional recognition, and facilitation, can help better integration of cross-disciplinary perspectives.

In this paper, we contribute:
\begin{itemize}
\item An empirical understanding of how early- and mid-career HCI faculty, in principal investigator roles,  experience multi-level challenges (individual, team, organization, research community, funding agency) in collaborating with health researchers throughout the stages of research: forming a team, conceptualizing research, implementing and translating research.  
% \item Empirical understanding of how early-career HCI faculty, in principal investigator roles, \textchange{ experience multi-level challenges (individual, team, organization, research community, funding agency)} in collaborating with health researchers throughout the stages of research: forming a team (institutional structures that support forming collaborations and challenges when those structures are missing, expectations for promotion that are misaligned with researchers collaboration goals), conceptualizing research (limited joint participation from different disciplinary perspectives, lack of understanding of researcher's roles, misaligned outcome expectations), implementing (misaligned timelines incompatible to human centered design research process, misunderstanding human centered design methods) and translating research (misaligned expectations for research outcomes).  
\item An introduction of a holistic Team Science lens, to identify challenges and support in collaborative cross-disciplinary teams that employ human-centered computing approaches across the research life-cycle 
% \item An understanding of the factors that improve or hinder integrating HCI perspectives with health research, including structures within universities, expectations of funding agencies, job structures, research community norms and networks, disciplinary approaches to research, and the invisible work of researchers. \textchange{todo understanding of the ways in which factors at a micro and macro level impact integrating HCI research with health research}
\item Recommendations and an agenda of future work for improving research integration and the successes of early researchers conducting cross-disciplinary research in HCI and Health, including  implications for individual researchers, universities, research communities, and funding agencies. 
\end{itemize}

% cut:
% These contributions help guide the CSCW research community towards an infrastructure that positively impacts and supports collaborative research in HCI and Health. This research outlines an agenda for HCI community members and leaders to support transformative, collaborative research through individual implications, as well as policy implications at the level of the research community, institution, and funding agencies. 

%% file: 02relatedwork.tex
\section{Related Work}
% \subsection{The field of Team Science}
Scientific initiatives are increasingly collaborative, with teams increasing in size and consisting of members from different fields who work together to solve complex problems \cite{national2015enhancing}. The field of Team Science has conducted decades-long efforts to understand and improve cross-disciplinary collaborative research and training programs to improve collaborative research \cite{national2015enhancing}. Team science involves ongoing research on conceptual frameworks to understand how to support more effective teamwork across disciplines. 
% To do so, models have emerged in team science that characterize the processes of successful collaboration in diverse teams \cite{salazar2012facilitating,hall2012four}.

Team science promotes a \textit{transdisciplinary} approach that requires team members to integrate their work in a way that transcends disciplinary silos \cite{national2015enhancing}. Transdisciplinarity \textit{“entails not only the integration of approaches but also the creation of fundamentally new conceptual frameworks, hypotheses, and research strategies that synthesize diverse approaches and ultimately extend beyond them to transcend preexisting disciplinary boundaries”} \cite{stokols2013transdisciplinary}. This goes beyond \textit{multidisciplinarity} approaches (teams working in sequential ways, independent from each other, with a goal to combine efforts and address a common problem), or \textit{interdisciplinarity} approaches (teams integrate concepts and methods from different fields and work to integrate differing perspectives, while still primarily staying anchored in their own field) \cite{national2015enhancing, stokols2008science}. 

% While teams might consist of members from multiple disciplines, they might not all work together in the same way. Different terms are used to describe cross-disciplinary teams depending on how well their work is integrated across disciplines. 
Transdisciplinary teams accelerate innovation and advances in scientific knowledge and address social problems \cite{national2015enhancing,spring2012emerging}. The field of Team Science \cite{national2015enhancing, stokols2008science} aims to better understand, evaluate, and support the effectiveness of science initiatives that promote research that draws on multiple disciplinary backgrounds and produces innovation across disciplines, particularly striving for supporting scientific collaborations of a transdisciplinary nature \cite{national2015enhancing}. 

%  Based on extensive research, the National Research Council proposes initiatives to support research teams at the level of the team, the institutions, the research field, or the funding agency level \cite{national2015enhancing}.

\subsection{Factors that Impact Research Integration}
A core aspect of doing transdisciplinary work is integrating knowledge across fields. Kraut et al.'s highlighted that research integration can be impacted by tasks, relationships, and the research climate \cite{kraut1987relationships}. The concept of \textit{research integration}  was introduced to CSCW and to Team Science through research by Balakrishnan, Kiesler et al that defined \textit{research integration} as \textit{"the extent to which a science team combines its distinct expertise and work into a unified whole"} \cite{balakrishnan2011research}.

\subsubsection{Team practices impact research integration}
CSCW research has highlighted that research integration is difficult to achieve due to workflows that are not aligned (e.g. communication of results between team members) \cite{balakrishnan2011research}, misaligned language \cite{olson2013working}, or lack of face-to-face interaction \cite{olson2000distance}. Salazar et al further expanded research integration in teamwork through a Team Science model highlighting the integrative capacity of teams as the team's ability to build communication practices, shared identity, and shared  conceptualization of a problem space to create integrated knowledge \cite{salazar2012facilitating}. Salazar highlights how social practices of integrating knowledge can impact teams to operate in constructive ways, adapt and adopt research differences to create new knowledge.  This could be due to adhering to discipline-specific norms of conducting science, claims of what is knowledge and who owns knowledge \cite{stokols2008science}, and insufficient training \cite{stokols2008science}.

To address such challenges, researchers have developed toolkits encouraging researchers to have conversations about goals, challenges, team member roles, and publication practices, to ensure the success of the team \cite{teamsciencetoolkit,bietz2012improving,teamsciencetoolkit-uw}. Further, Team Science proposes the development of training programs through which young researchers and students get trained in multiple disciplines \cite{national2015enhancing}.

% \textmodified{\paragraph{Aligning methods and goals}}

% These toolkits are either designed for a general scientific audience, or modelling practices and norms in medical schools or health related department.

%  \color{purple}

\subsubsection{Infrastructure impacts research integration}
Multiple organizational factors can support collaborative teamwork such as credit and rewards, and funding \cite{stokols2008science,toubia2006idea}. Individuals engaged in creative work can have a hard time doing that in the organizations they belong to because they need to balance the challenges of demonstrating the output of the creative work and navigating the formal structures that coordinate the work they do \cite{adler2011combining}. This can be particularly difficult within university systems, where faculty behavior can be impacted by evaluation and reward systems, workload allocation, professional development opportunities, and leadership \cite{national2015enhancing}. 

% To incentivize change, it is most effective when organizations take both a "top-down" and a "bottom-up" approach [ref austin 2011]. 

\paragraph{Funding Incentives.} Institutions have created funding and organizational structures to promote cross-disciplinary research. Funding agencies have launched programs to promote translational science and collaboration across disciplines \cite{nsf-sch,zerhouni2005translational}. Universities also create seed grants for cross-disciplinary collaborations \cite{national2015enhancing}. Organizations  (e.g. funding agencies) that created networks of scientists from different disciplines have made it easier for researchers to start cross-disciplinary research \cite{national2015enhancing,nci-pathways}.

\paragraph{Promotion practices.} The US National Academies of Sciences, Engineering, Institute of Medicine, have ranked promotion and tenure as one of the top impediments to cross-disciplinary research \cite{engineering2004facilitating}. A top reason for that is the uneven evaluation of individual contributions to cross-disciplinary research. For example, in co-authored papers, credit might not be awarded fairly between team members \cite{merton1968matthew,merton1988matthew}, and author order might vary between fields and countries \cite{tscharntke2007author}, which can affect career advancement. To address such issues institutions have come up with guidelines for how to evaluate cross-disciplinary research, through promotion guidelines \cite{hall2013recognition}. Over 32 institutions that promoted cross-disciplinary research took such measures \cite{hall2013recognition}. Although initially, such guidelines were minor, there are increasing resources and toolkits for language that can be used in promotion packages and in presenting contributions of cross-disciplinary work \cite{teamsciencetoolkit-uw-promotion}. 

\paragraph{Institutional structures to form cross-disciplinary teams.}
Universities have launched cross-disciplinary centers across the country in fields such as sustainability, biodesign, and social change \cite{frodeman2017oxford,repko2020interdisciplinary,glied2007institutional}. However, assembling successful teams across disciplines can be difficult, and might favor some researchers with a history of collaborations \cite{hall2018science}, more experience \cite{hall2018science}, and more social capital \cite{li2013co}. The approaches used to find collaborations are often unstructured, relying on the institutional proximity of research networks \cite{hall2018science}. To facilitate collaborations across disciplines, research universities have promoted research networking systems, though their effectiveness is not known yet \cite{kahlon2014use}. Given the difficulties in assembling a team, there might be further challenges to new investigators who are less experienced and are joining institutions new to them. 

Researchers have identified many challenges in conducting cross-disciplinary research in a variety of fields, at the team and infrastructural levels. But there is limited understanding of the challenges that researchers in the field of HCI encounter in using HCI methods in other fields. Team Science brings a team level and infrastructure level focus on collaborations and proposes tools that address research integration at the level of the team members and initiatives that can support collaboration at institutional levels. To understand the complexities of utilizing HCI methods across disciplines, there is a need to understand how team and infrastructure level challenges reflect in HCI collaborative teams.

% It is still unclear how these toolkits could help HCI researchers, or how they should be adapted to HCI researcher needs. This demonstrates a need to understand the 

% Despite this extensive knowledge, we have less insight into how Team Science concepts can be used to identify and understand challenges that occur in transdisciplinary CSCW research. In this paper we draw on team science models, described below, to understand how collaborations are formed and how health and HCI knowledge is integrated into collaborative work at the intersection of HCI and health.

\color{black}

\subsection{The research lifecycle: Hall's four-phase model}

% Transdisciplinary work can lead to higher innovation, to solve complex problems, but there are numerous challenges that come in the way of doing such work, such as diverse goals of team members, unstructured problems, different methodologies, and many more. 

Knowledge integration can be difficult to achieve throughout the different stages of the research process. In this study, we draw on Hall's four-phase model to describe challenges in integrating team members and perspectives  across different stages of a research project timeline. The four stages of research identified through a team science lens involve collaboration from the formation of a team and definition of a project, to the translation of the research into the real world. Research teams have different goals at different stages of research, might encounter different challenges and need different types of support. Therefore it is important to understand the challenges that teams encounter at different stages of research \cite{national2015enhancing}. By understanding challenges at different research stages we can develop targeted interventions, or understand challenges in collaboration at different levels.  

\paragraph{Development phase}
In a transdisciplinary model, the first stage of research is the \textit{development stage}, where research teams are assembled and define the problem space that will be researched. At this stage, the team articulates the complexities of the problem space and identifies the relevant research disciplines, including a potential group of collaborators \cite{hall2012four}.  The team-assembling step is critical to the success of a team's ability to generate transformative innovations \cite{hackman2012causes}.  

\paragraph{Conceptualization phase}
During the \textit{conceptualization phase}, researchers develop novel research questions, hypotheses, frameworks, and a research design that integrates the disciplinary perspectives of the different collaborators \cite{hall2012four}. However, it is difficult to develop a shared problem conceptualization and develop shared goals \cite{stokols2008science,national2015enhancing}. Team members come from a different disciplinary background with differing goals (e.g., to make theoretical vs. applied contributions) \cite{stokols2008science}, or differing needs for career advancement \cite{national2015enhancing}. When teams do not address such misalignments, it can lead to the emergence of subgroups within a team that can be divisive or hinders progress towards the project goal \cite{salazar2012facilitating, campbell2005overcoming}.

\paragraph{Implementation phase}
During the \textit{implementation phase} research teams launch and conduct transdisciplinary research \cite{hall2012four}. During this stage, researchers collaborate to refine and extend the research questions and methods used. At this stage, team members are faced with the  challenge of integrating diverse knowledge and methods \cite{salazar2012facilitating,national2015enhancing}. Without sufficient overlap in how a problem is conceptualized through research questions or standards of conducting research, it can be difficult to apply knowledge from multiple disciplines and generate a high-quality solution \cite{salazar2012facilitating,fiore2004process}.

% However, less is known about how researchers navigate such methodological challenges to maintain successful collaborations with colleagues outside of the HCI and CSCW fields. 

\paragraph{Translation phase}
During the \textit{translation phase}, researchers create pathways to apply their findings to real-world problems \cite{hall2012four}. For this stage researchers identify translational partners to help initiate community outreach activities and to plan and define translational goals. Depending on the research this can involve, for example, translating basic science to research that involves clinical studies or translating clinical studies to public health programs and policies. 

% \color{brown}

\subsection{Knowledge gaps in HCI-Health research collaborations}

Health-HCI cross-disciplinary teams encounter many challenges in conducting research. The challenges identified are typically experiential reflections on research by individuals, as described below. 

At a methodology level, HCI and health researchers have challenges with shared conceptualizing of problems  through the contrasts of needing to balance uncovering the needs of users with the goals of health researchers to develop interventions, theoretical frameworks, and evidence-driven practice \cite{blandford2018seven,blandford2018lessons}. Health researchers might aim to conduct longitudinal deployments and evaluate the effectiveness of interventions that use legacy technologies like text messaging, an incompatible approach with HCI's desire for novelty \cite{mamykina2021scaling,cibrian2022interdisciplinary}. Different evaluation of technology can also lead to misaligned timelines \cite{cibrian2022interdisciplinary}. HCI researchers might instead innovate technology and evaluate prototypes \cite{mamykina2021scaling,klasnja2011evaluate,blandford2018seven}. Research outcomes between HCI and Health are different, which suggests a need to choose to work on problems that involve the entire spectrum of understanding user needs and clinical outcomes \cite{blandford2018lessons,blandford2019hci4health}. When it comes to translational work, in HCI translating work to communities has sometimes been done as part of action research \cite{hayes2011relationship}, by involving community partners in the research, developing it together, and having a socially meaningful impact. However, it is unclear how HCI researchers engage and experience collaboration for translation purposes.  

Methodological differences in HCI and Health can lead to challenges in executing research. HCI emphasizes the user needs, but that can lead to designing more complex health interventions that are difficult to evaluate \cite{mamykina2021scaling}. On the other hand, working in a clinical setting and building relationships with clinicians can be constrained due to complex hospital infrastructure \cite{aggarwal2020lessons,blandford2015strategies, furniss2014fieldwork}. There can be limited time with patients and clinicians, which can lead to inappropriate solutions \cite{buis2020common,williams2020hci}. Health interventions can be predisposed to risks of equity imbalances \cite{veinot2019leveling}. Researchers have to navigate ambiguous and varying IRB requirements to collect digital health data for their studies \cite{huh2020s}. Researchers have had to adjust methods (e.g. wizard-of-oz) to conduct research in the health setting \cite{mitchell2021curtain}, and consider logistical considerations of deploying technologies \cite{kulp2018design}. Researchers also are required to strike a balance between anonymizing participant health information and sharing enough of this information for their work to be rigorous and reproducible \cite{abbott2019local}, or to have appropriate compensation for participants depending on the regulations of the health system \cite{pater2021standardizing}. Although we know about tensions that  HCI and Health research encounter at a methodological level, there is limited understanding of how researchers encounter such challenges in collaborations, their experiences, and what they do to navigate them. 

At an infrastructure level, HCI researchers have discussed challenges in working with clinicians \cite{aggarwal2020lessons,furniss2014fieldwork,blandford2015strategies}, however, there is little empirical knowledge of how HCI researchers go about forming these collaborations to develop research projects that address complex problems.

% In summary, past evidence shows the practical challenges of conducting HCI research in the health domain, and provides insights into the practices researchers follow when applying human-centered design. However, in considering the body of research about researchers at the intersection of HCI and health, there is a gap in our understanding of what challenges HCI researchers face in establishing such projects through collaborations with health researchers. We know little about how early-career HCI researchers navigate or cope with these challenges, and what impact these challenges have on their career goals. 

In summary, we use Team Science as a lens to understand the challenges that HCI researchers encounter when working with health researchers, throughout the different stages of the research process. We present findings organized based on Hall's model of research stages and highlight the challenges that HCI researchers have in integrating perspectives in collaborative work between health and HCI researchers. We highlight the team level and infrastructure level challenges as they manifest throughout the research lifecycle.

\color{black}

%% file: 03methods.tex
\section{Methods }

We conducted semi-structured interviews with HCI researchers who had expertise in technology design to understand the challenges they face when collaborating with health researchers. This study was deemed exempt by the authors' Institutional Review Boards.

\subsection{Semi-Structured Interview Procedure}
To gain a deep understanding of participant experiences and challenges, we conducted semi-structured interviews. Each participant was asked about their experience working on collaborations with health researchers, what challenges they faced during these collaborations, and what lessons they learned from their experience. The interview protocol evolved in response to interviewee responses. We conducted two pilot interviews not included in the data presented here. The initial interview protocol focused on the challenges of conducting a collaboration and in using human-centered methodology when the collaboration was already established. The interviewees surfaced challenges in finding and forming collaborations as well as infrastructure challenges. Based on the two pilot interviews we conducted a more in-depth literature review in Team Science, which led us to add questions related to experiences participants encountered at different levels: individual (e.g. how  collaborators were chosen), team (e.g. roles in the project), methodological (e.g. use of design methods through the project life-cycle) and organizational level (e.g. what helps in finding collaborators). 

All interviews were conducted remotely over Zoom. A maximum of two authors were present for each interview. One author was responsible for conducting the interview with the participant. A second author, when present, was responsible for asking follow-up questions and taking notes on key details. Interviews lasted between 40-70 minutes, depending on the availability of the participant, and were audio-recorded and transcribed for thematic analysis. After completing their interview, participants were offered a \$30 gift card as compensation. 

\subsection{Eligibility and Recruitment}
To recruit eligible participants, we began by identifying eligible researchers in the CHI conference proceedings within the last 3 years (2019-2021) who had published a paper on health or wellness and used design methods. We identified papers that mentioned the topic of health, wellness, mental health, exercise, or diet in the title or abstract of the paper. We corroborated that the papers were broadly on the health topic rather than something completely different. We emailed people who seemed to be faculty based on their public profile, in either HCI fields or Health fields. We did not want to make any assessments about who identified to be an HCI researcher or not. Using contact information available in the public domain (e.g., websites and online CVs), we verified potential interviewees' career stages and contacted them over email. We emailed one or two of the faculty last authors of each paper selected, excluding repetitions. As we conducted interviews, participants also contributed names of potential participants we could reach to for an interview. We emailed a total of 78 people, and 28 replied being interested to be interviewed. Of the 28 potential participants, 17 identified as primarily an HCI researchers, and 11 identified as primarily as health researchers. The 17 participants who identified as primarily HCI researchers constituted the participant sample we describe in this paper. \textadd{Of the 17 participants, 14 were identified through CHI proceedings and three through snowball sampling (participants who had not published at CHI during the determined period). We targeted individuals at different institutions. If a participant from a given institution confirmed an interview, we generally tried to sample from a different institution to obtain more diversity in participants. We also posted announcements of recruitment through social media, but they did not result in additional participants.  We continued participant recruitment until we reached data saturation, which occurred at participant 15. Because we had already scheduled two more participants, we completed their interviews as well. We had multiple participants who were available, but were not interviewed due to having reached data saturation. }

% Over email, we described the goals and procedures of the study, assessed their interest in participating, and confirmed their eligibility. 
% todo say how we selected HCI researchers

HCI researchers were considered eligible for participation if they were: (1) at least 18 years of age, (2) involved in technology projects that involved both health researchers and technology/design researchers, (3) worked in the United States, and (4) were identified as early-career or recently tenured (in the past 3 years) at an academic institution. We defined health researchers as collaborators who had formal training in health or clinical research and may also be  healthcare providers or in a similar field (e.g., clinical psychology). We did not include researchers who might do work at the intersection of HCI and Health, but might not employ human-centric design processes (e.g., researchers researching algorithms or hardware were not included).

We focused on United States-based researchers. The United States' funding infrastructure and the complexities of its healthcare system are distinct from those of other countries. We also focused our recruitment on early- and mid-career researchers because we wanted to capture the full life cycle of a collaboration that early-career researchers experience, and how the participant's approach toward navigating challenging collaborations may change as they moved to the next stage in their career.
% todo say that we focused on the infrastructure in the US

\subsection{Anonymization}
Due to our study's focus on researchers within a subset of the HCI and Health community, we took several steps to anonymize and protect the identity of our participants using norms similar to other studies done with participants from the HCI research community \cite{liang2021embracing}. Most participants revealed vulnerable information or information that they requested was anonymized about their collaborators, especially involving more senior collaborators. To prepare our interview transcripts for data analysis, we excluded information such as institution names, previous professional roles, and other potentially identifiable details. Throughout this paper, we present results in aggregate whenever possible. We only present participant backgrounds at an aggregate level because associating such details with participant IDs might risk deanonymization. We redacted the names of research projects or specific health topics that participants studied because in certain health application domains, some of the participants might have been the only person in the HCI community who was early-career and working on that topic. The summary of research domains and context of participants is broadly defined because in each given domain, there might be very few people working on a particular topic with a cross-disciplinary perspective as one of our participant sample. 

\subsection{Participant Background}

We enrolled a total of 17 participants in our study from 16 different universities in the US. Our sampling consisted of 11 Assistant Professors (7 were in their first 3 years as faculty, 4 were in year 4 or later) and 6 Tenured (but not Full) Professors within the 3 years prior to the interview. In addition, 14 participants came from research-intensive (R1) universities and 10 of those participants had a medical school within their institution; 1 participant was at an R2 university with a medical school, and 2 participants were at a Liberal Arts College with no medical school. Their self-reported educational backgrounds included degrees in behavioral health, clinical health, technology, and design. All participants had a degree and technology or design. All participants had publications in the HCI field, particularly given our criteria of recruiting from the largest HCI conference, the Conference on Human Factors in Computing Systems (CHI).

% \textchange{We aimed to recruit participants who identified as a using human centered design methods, and how did not identify as being a medical or clinical researcher.}
% We recuirted participants with a range of involvment in health projects, including some participants who were new
At the time of their interviews, the majority of participants (76\%) participated in less than 10 projects that involved both health and technology/design researchers. These participants' experience conducting research projects in the health domain ranged from 1 to 10 years (12 participants), to 10-20 years or more (5 participants). Participants reported that their work involved cross-disciplinary collaboration 20-40\% of the time (3 participants), 40-60\% of the time (1 participant), 60-80\% of the time (9 participants), and 80-100\% (4 participants). Three of the participants held (or had in the past) a primary appointment in a school of medicine, Biomedical Informatics, or Public Health. Most participants published their research in health publication venues as well as HCI or technology venues. We analyzed the public Google scholar profiles of the participants at the time of the interview and counted how many publications (including papers and abstracts) they had published in what appeared to be health journals and conferences (e.g. AMIA, JAMIA, JMIR journals, specialized journal such as psychiatry, nutrition, exercise, clinical medicine, rehabilitation, medical informatics, etc). Participants had published in health journals or conferences: over 30 times (2 participants), 11-20 times (3 participants), 6-10 times (3 participants), 2-5 times (7 participants), never (2 participants). These numbers were collected across the entire publication career of participants, and are impacted by their seniority level (participants with most publications were generally more senior).

The diversity of engagement in cross-disciplinary research enabled us to gain an understanding of the challenges that different types of scholars encountered in doing cross-disciplinary work, including scholars at different types of institutions (R1, R2, Liberal Arts), at a range of stages of their career -- early to mid-career, with a range of primary appointments in computing departments and in medical schools. 

Participants ranged in age from 31 - 50 years old (mean = 37 years, median = 36 years). One participant did not report their age. Our sample also allowed participants to self-identify their gender, which they reported as woman (15), non-binary (1), and man (1). A total of 11 participants identified their race as White/Caucasian, 5 identified as Asian, and one was not sure how to identify their race.

\subsection{Data Analysis}
We used deductive and inductive analysis to analyze all the interviews. After each interview, the authors wrote memos and summarized important themes. For the deductive approach, we created codes driven by Team Science literature \cite{national2015enhancing} about the challenges team members typically encounter (e.g. individual's research and professional goals, team goals, roles in the team, methods challenges, institutional infrastructure such as training programs, departmental structures, research infrastructure such as mentor networks, promotion expectations, etc). Our inductive analysis relied on a line-by-line descriptive coding approach \cite{saldana2021coding}. Example codes that emerged through inductive analysis were: 
working independent of collaborator, training or being trained about methods, expectations about what is data, participation in project on-demand. Both authors then open-coded two interviews together to begin developing a codebook based on the memos. After creating a preliminary codebook, the two authors coded a new set of transcripts independently, then discussed coding patterns and resolved coding discrepancies. The codebook was revised and applied to all the transcripts by at least one author. Throughout this process, all authors met several times per week to discuss, refine and conceptualize themes. \textadd{The deductive analysis led to themes that related to challenges that cross disciplinary teams encounter according to Team Science literature \cite{national2015enhancing} and across different stages of research \cite{hall2012four}. For example, methodology, goals, or outcome misalignments are common challenges experienced in teams. Such challenges are broad, thus the data helped us identify sub-themes of how such situations were experienced by participants in our sample.}

\subsection{Positionality of Authors}
All authors identify as early-career researchers. Two had seven years of experience working on human-centered computing studies and collaborating with researchers working in the health domain. One author was a student, has no experience working with health researchers and provided a newcomer perspective to the study and data. All authors also have been students and/or held early-career academic roles at R1 institutions, a privilege that influences the makeup of their professional networks, and potentially the visibility and willingness to participate in this study for eligible participants. The senior authors of the paper were motivated in part by their experiences and observations working in teams with health researchers. They were driven to articulate the invisible challenges that early-career faculty face when working in such teams, and to learn from other researchers' approaches to navigating cross-disciplinary research in HCI and Health. 

%% file: 04results.tex
\section{Results}

In this section, we detail the challenges and factors that participants identified as impacting their ability to achieve successful collaborations throughout the different stages of research: development, conceptualization, implementation, and translation (Table 1). We present the HCI researchers' experiences in  building collaborations, aligning goals and outcomes, and aligning disciplinary approaches. 

Participants encountered support and barriers in creating partnerships and integrating perspectives at multiple levels: the institutions they were part of, funding agencies, their job structure, the norms in the research field, research communities, research practices. Some of these structures made it easier to find collaborators, align goals and methods. However, when researchers encountered barriers it impacted how integrated the  researcher's work was, the extent to which their research perspective was included in the work, the speed at which they could make progress, the overall impact of their research contributions, and career advancement they could achieve. For brevity throughout the remainder of this paper, we will refer to the  participants as \textit{HCI researchers}. We will refer to the collaborators of the participants as  \textit{health (research) collaborators}.

\subsection{Challenges in integrating perspectives across all research stages}

Participants mentioned several challenges that impacted their ability to engage with health researchers closely throughout the research process. These reflected a lack of experience and training in how to work with health collaborators, as well as structural aspects of the health researcher's job structure that impacted their involvement in research projects. 

% \textchange{\textbf{Individual factors }}
\subsubsection{Lack of experience in starting and maintaining collaborations with health researchers}
Several participants mentioned they did not have prior experience building collaborations, because they did not need to set up their own collaborations during their PhDs or postdoc because their mentors did so (C3, C8, C10, C14, C15). They had a period of transition where they had to learn  how to form and lead their own collaborations. C7 explained that they were not used to building a longitudinal relationship with a collaborator, but rather they had treated their collaborator as a means to finish the PhD: \textit{ "I had a very narrow vision, what health collaborators roles are... just building that missing piece... which is the clinical domain knowledge... but really, if it's going to be a prolonged collaboration, that just doesn't work, because people's motivation might be different. Some people might enjoy that, but not many, especially when, they are super busy"}. C7 mentioned making a similar assumption early in their faculty years when they regretted not involving their collaborator more: \textit{“we wrote a paper... out of that project, but... he didn't end up becoming a co-author. He was okay with that. But after that, there was no follow-up work... you do need access to patients in your work. But if your collaborator doesn't have a clear contribution intellectually, except for like giving access to their patients... it's a little bit tricky whether to invite the person or not, when it's time to write a paper”} (C7).

% \textchange{\textbf{Organizational factors}}
\subsubsection{Lack of training opportunities to work in the health domain}
Several participants (C5, C17) mentioned that they wished they had faculty training in working in the Health field, and with health researchers. They wished they could apply for a training grant funded by NIH in which they could get training in Health research, along with getting mentorship from health faculty. Such training grants are available through NIH for people who gained a PhD, working on relevant health problems, but are primarily targeted to health researchers. However, HCI researchers thought this would be disadvantageous to their position as an assistant professor. C17 thought that an NIH training grant would require them to teach less, which could negatively impact tenure expectations in a computing or engineering department: \textit{"K award has disadvantages for junior faculty, because... [it] takes away from some of your teaching, and you want to sort of show when you go up for tenure"}.

% \textchange{\textbf{Organizational and Funding approaches factors}}

\subsubsection{Differences in job structure: difficulties in securing time}
Some HCI researchers experienced incompatibilities with their collaborators because their collaborator's job did not match the expected engagement or contribution (C3, C5, C16) . Some of the participants worked with collaborators whose salaries were partially determined by grant funding (C2, C3, C5), HCI researchers felt pressure to secure their collaborator's participation. C5 found it necessary to raise funding, so they can support their collaborator's time, particularly for people who have a different job that takes most of their time: \textit{"what I learned from this collaboration is that I should not have a clinical collaborator who doesn't have any time, the funded time devoted to research. So he has... 100\% hospital [time]. And so he's doing this as a side project"} (C5). Participant C5's experience was confirmed repeatedly when a different potential collaborator became less involved because the collaborator received a different grant to fund their salary. 

Another challenge was the different ways HCI researchers needed to work around the limited time that health researchers had, and maintained a relationship that was respectful of limited time (C1, C7, C13, C17). This was particularly the case when the health researcher also had clinical care duties, in addition to their time allocated to research.  C17 posed this as a challenge:\textit{ "how do we bring them [health researchers] in, but respect their time and not be burdensome?... we don't want to be extractive... with the little chunks of time they have... they don't really have the capacity to be full... we can't ask them to spend hours a day... on the research. And that means that we're doing research when they're not updated [about]... if we're only speaking with them an hour a week"}. HCI researchers navigated this challenge by engaging with the health researchers based on their availability, such as having meetings at times of day that might have been inconvenient (C7, C13), because they found the contributions of the collaborator so important.
% \textit{"this clinician is just so open to HCI. And every time you're talking about certain tools or different approaches, he is just so intrigued about different ideas and just being available. He is very busy"} (C7). 

\setlength{\tabcolsep}{10pt} % for the horizontal padding
\begin{table}[ht] %[ht] [hbt!]
    % \color{purple}

    \caption{Factors that Impact Collaboration Between HCI and Health Researchers at Different Stages of Research.}
    \scalebox{0.9}{%
    \label{tab:factorstransdisciplinarity}
    \begin{tabular}{p{ 0.30 \linewidth } p{0.2 \linewidth} p{0.5\linewidth} }
    \hline
    \textbf{Stage of Research} & \textbf{Micro and Macro- levels of impact} & \textbf{Factors that Impact HCI-Health Collaborations}  \vspace{5pt} \\ \hline 

    \multirow{4}{ \linewidth }{\textbf{Developing Research \& Starting Collaborations} \newline Researchers come together from different disciplines and define the problem space that will be studied} & 
    Institutional & 
    Postdoctoral training programs, university level grants and cross-disciplinary meetups, department structures promoting cross disciplinarity   \vspace{5pt} \newline
    Expectations for tenure and promotion \vspace{5pt} \\
    & Funding agency & Funding agency-level requirements and initiatives \newline \\
    & Research field & Networks of mentors, peers and collaborators \vspace{5pt} \\
    & Individual factors & Research values, needs to developing research agenda and network
     \vspace{5pt}  \\
    \hline
    
    \multirow{2}{ \linewidth }{\textbf{Conceptualizing Research} \newline  Researchers develop novel research questions, hypotheses, frameworks, and a research design that integrates across the disciplinary perspectives of the different collaborators } &
    Team level collaboration &
    Joint conceptualization of research \vspace{5pt} \newline 
    Limited understanding in the team of HCI researcher role \vspace{5pt} \newline
    Ambiguity about the future of the research project \newline \\

    & Funding agency  &
    Funding agency expectations of researcher roles and research outcomes \newline  \\
    % \vspace{5pt} 
    \hline

    % \textbf{Conceptualizing Research } \newline 
    % Researchers develop novel research questions, hypotheses, frameworks, and a research design that integrates across the disciplinary perspectives of the different collaborators &
    % Team level collaboration &
    % Joint conceptualization of research \vspace{5pt} \newline
    % Limited understanding in the team of HCI researcher role \vspace{5pt} \newline
    % Ambiguity about the future of the research project \vspace{5pt} \newline
    % Funding agency expectations of researcher roles and research outcomes \vspace{5pt} \newline
    % % Researcher's comfort in voicing research perspective based on research seniority 
    % \vspace{5pt} \\
    % \hline
    
    \textbf{Implementing Research}  \newline
    Researchers collaborate to refine and extend research questions and methods used, and execute the research &
    % Funding agency \vspace{5pt} \newline
    Training: Methods &
    Different timelines for executing research \vspace{5pt} \newline
    Understanding of HCI research and methods: \newline
    - Skepticism of qualitative research and design methods \newline 
    - Misalignments of what is considered data \newline
    - Centering user needs in conflict with clinical expertise \newline
    - Artifacts facilitating knowledge integration \newline
    - Differences in terminology \newline
    % Approach to conduct research outside of collaboration boundary
    % \vspace{5pt}
    \\
    \hline
    
    % \textbf{Translating Research } \newline
    % Researchers create pathways to apply their findings to real-world problems &
    % Training: Methods \& Collaboration approach   \vspace{5pt} \newline
    % Research field \& Institutional  &
    % Level of expertise in translational approaches (e.g. clinical trials)  \vspace{5pt} \newline
    % Tensions in producing technological innovation vs practical translation of research \vspace{5pt}  \\
    % \hline

    \multirow{2}{ \linewidth }{\textbf{Translating Research } \newline Researchers create pathways to apply their findings to real-world problems} &
    Training: Methods \& Collaboration approach &
    Level of expertise in translational approaches (e.g. clinical trials)  \vspace{5pt} \newline \\

    & Research field \& Institutional  &
    Tensions in producing technological innovation vs practical translation of research \vspace{5pt}  \\
    \hline
    
    Across all stages &  
    Training & 
    Limited experience to start and maintain health collaboration \vspace{5pt}  \newline
    Limited training to grow Health skills \vspace{5pt} \\
    
    \color{white}..\color{black} &  
    Institutional & 
    Differences in job structure and time commitments of HCI and Health collaborators\\
  
    \end{tabular}
    % \vspace{-13pt}
    }
    \color{black} 

\end{table}

\subsection{Developing research: factors that impact starting collaborations}

In the research development stage, researchers come together from different disciplines and define the problem space that will be studied \cite{hall2018science}. Below we identify several factors that impacted whether HCI and health researchers started collaborations together: institutional supports and programs, funding source requirements, job structure, or individual efforts. These structures impacted if researchers could come together to start shaping the research. 

% \textchange{\textbf{Organizational factors}}

\subsubsection{University structures: barriers and facilitators to starting collaborations}

Participants identified several institutional structures that facilitated their finding a collaboration at the level of universities and funding agencies: training programs, cross-disciplinary centers, funding opportunities, or cross-disciplinary workshops. Such opportunities were selectively available or not at different institutions and sometimes impacted HCI researchers in different ways. 

\paragraph{Postdoctoral training programs} 
Several HCI researchers had done postdoctoral programs (6 mentioned having done one). All participants mentioned that during their postdoctoral programs they started or joined new projects. This created collaborations with health researchers that several of them continued after the postdoc. One participant was part of a training center that brought HCI and health researchers together during their postdoctoral program. During this experience, they trained along with health researchers to closely work together with health researchers and use design practices, which enabled them to continue working together after the completion of the postdoctoral program. Despite finding collaborators during postdocs, HCI researchers moved to a new institution as assistant professors. Therefore they had to restart some of their process of finding collaborators when starting their professor positions.   

\paragraph{University level grants and meetups}
Universities that included programs to facilitate collaborations between researchers allowed participants to find other collaborators, through opportunities for funding such as seed grants and networking events. HCI researchers had an easier time finding collaborators when the institutions they were working at that included formal structures that facilitated forming cross-disciplinary collaborations (C2, C5, C13, C17): \textit{“They sometimes host like meetups for... researchers from different fields to kind of talk about what they do... There are... internal funding”} (C13). 

\paragraph{University structures and departments}
Participants had an easier time finding collaborators at schools that had research centers that included cross-disciplinary collaborations as part of their mission, and universities that had medical schools. Others encountered barriers in finding collaborators when such structures were lacking (C4), for example in \textit{"rural"} locations with less infrastructure (C6).

The institutional context and the type of departments in a university affected how HCI researchers were perceived.  For example, C3 needed to understand the type of design scholarship that their collaborator had experienced, because of the collaborator's experiences across different departments in the university: \textit{ "the collaborators that I'm dealing with... at my university are used to either dealing with communications or with the [art and design division]... they're... not used to the particular flavor... specifically, the art and design folks are not interested in getting research publications in the same way... the communication scholars... they are much more interested in...  evaluating something in use, then sort of this sort of building, collaborating on the design piece" } (C3). 

% \textchange{\textbf{Funding agency}}

\subsubsection{Funding agency-level requirements and initiatives}
A factor that impacted searching for collaborators was the need to have a collaborator to submit a grant with. Depending on the funding agency, the HCI researcher might search for a collaborator, or the health researcher might seek an HCI collaborator to fill in grant requirements. This sometimes occurred on short notice, to complete a grant deadline. Such short notice engagement sometimes led to successful collaborations (C2, C16), but not always (C16, C17). C6 experienced the requirement to find a collaborator and despite repeated efforts, they could not find one: \textit{"I submitted a grant. And I know one critique that I didn't have clinicians at collaborators...  maybe just write a support letter for me... I was actually planning to submit [grant] and I give up, because... I haven't found a clinician as a collaborator yet... this definitely caused a negative impact on my career"} (C6). 

At times, funding agencies organized workshops that were helpful in finding collaborators. C5 found a collaborator at professional workshops, invitation-based, organized by a foundation (C5). 

% \textchange{\textbf{Research community factors}}

\subsubsection{Research networks of mentors, peers, and former collaborators}
HCI researchers relied on mentor networks to find collaborators. For example, participants who had been part of postdoctoral programs, had joined projects where collaborations were already established. Alternatively, their mentor helped them connect to health collaborators. 

\paragraph{Mentors facilitating connections}
When HCI researchers did not have a formal mentorship structure, they relied on mentors at their current institution or from their former institutions to help them connect with relevant collaborators (e.g. prior mentors and advisers, department colleagues, dean of the school) (C4, C6, C10). Some participants relied on more senior faculty to make such connections and highlighted the importance of having senior researchers who built infrastructure at their university for conducting cross-disciplinary research (C11, C17). C4 also found it helpful that the collaborator was willing to act as mentors: \textit{"the willingness of the collaborators to kind of like mentor other people that are getting in the space seems to be in my case has helped me tremendously... dedicating time in things that might seem trivial, but for me were quite new"} (C4). In other cases a mentor or other third party recommended the HCI researcher to a team of health researchers who were seeking collaborators (C3, C7, C16). 

Although most participants in our sample found collaborators that appeared suitable for initial collaborations, participants viewed a lack of collaboration as hindering their career advancement. For example, C6 struggled to find a collaborator for grant applications through health researchers and several third parties (mentors, colleagues, dean), for over a year.  

\paragraph{Serendipitous opportunities to find collaborators}
Participants who had prior collaborators reached out to those for getting the help they needed (C8, C12). However, other researchers were at institutions where they did not have other health researchers to facilitate such connections (C4, C6, C8, C9, C15). Researchers tried to leverage research networks also through giving talks at their own institutions and others (e.g. in courses taught by potential collaborators, at other institutions) (C5, C17), or invited potential collaborators to give presentations (e.g. in the classes they taught) (C17). Other facilitators of collaboration included having similar goals and being at similar stages in their career with health researchers, such as being at the same early stage in their career (C4, C12), or being able to relate with similar life milestones in their personal lives (C3). 

% \textchange{\textbf{Promotion expectations, Organizational, Research community}}

\subsubsection{Expectations for promotion and career advancement}   
Some HCI researchers decided that they would not engage in research that did not directly contribute to their primary research agenda, to focus on what they contribute and produce to achieve their goal of tenure: \textit{"I want to try as an assistant professor as a pre-tenured faculty... I want everything that I do to contribute to that... in my experience, many of us in this space, have had to kind of grapple with the well, I could do that. I could be part of that project... But... that's not going to help me... I really am [focused on tenure]... I need to be really... selective" }(C16). While tenure priorities restricted what projects C16 would pick, at times it also gave them an opportunity to exit projects that were not going as expected, by using the necessity to focus as a reason to exit the collaboration. 

C3 also felt concerned that having too many projects in which their role is one of being a "consultant" for the health researcher, would not support their tenure case: \textit{"If I had too many projects at once that were sort of those service consulting kind of projects…. That's when folks would start to get worried... because then I'm just sort of like, work for hire. Rather than contributing to my own research"}. However, C3 thought that the strategy of doing projects that were not core to their research helped them find collaborations for their primary work: \textit{"I need to pursue this strategy... to cultivate these relationships in order to build these collaborations for my own core human-computer interaction"}. 

% \textchange{\textbf{Individual factors}}

\subsubsection{Individual factors that impact starting a collaboration}
Participants had developed criteria for what type of collaboration they would join, and how they would decide about it. However, HCI researchers we interviewed were junior, therefore they had a need to develop research agendas and relationships to start their research quickly. Sometimes researchers joined collaborations that were non a fit research-wise but would provide them with other benefits of career development. Participants took into account promotion requirements when making such decisions. 

\paragraph{Collaborator's research values}
When deciding if a health researcher is a suitable collaborator, HCI researchers sought to understand the researcher's worldview. Participants wanted to know if the collaborator adopted a user-centric perspective or a patient-centric perspective (C1, C10, C17). For example, C1 talked about how her research approach and values aligned with their collaborator: \textit{"I think we are more aligned in terms of our approaches where it's more inductive, it's more qualitative... she's actually way more fundamentally oriented towards understanding people first than I am even"}. A lack of this patient-centered worldview was noted as an issue and potential barrier to collaboration, for example: \textit{"red flags for me are people who who aren't deferential to the patients"} (C10). Participant C17 also  added that they valued shared decision making, how patients are involved, timeline coordination, and novelty of the research: \textit{"Do we have the same values here?...  what it will really mean to share decision making, what's the orientation toward society towards patients, toward other subject matter experts however we're bringing them in, do we have time to do the right work... if it's not going to lead to something novel, I might still say no"}. 

%  So someone where like, I've just met them, and they become a participant before they become a collaborator, because I haven't figured out how to make a collaborator yet. But they're like a researcher and they could potentially be a collaborator, but I like invite them to just sit in an interview to get feedback for my student, which, which is not that different from like a meeting, I asked them for a meeting to give me advice on my research or to give my students advice, but we just like, turn on the recording, and use that as data. (C2)

% \textchange{\textbf{Research community and promotion expectations}}

\paragraph{Needing to develop their research network and research agenda}
Enrolling in a collaboration depended on the needs that HCI researchers had at the current stage in their career. This emerged by needing to develop a research agenda and learn new skills such as grant writing. Participants sometimes chose to pursue opportunistic collaboration to which they were invited, based on the alignment of the health researcher's work with their own research interests as well as with other career goals. In some cases, the HCI researcher was not looking for the specific collaboration to occur (e.g. the specific collaborator or topic), but when a health researcher asked them to join their project they saw it as an opportunity to advance their network and learning goals. Some researchers wanted to engage in research within a new domain or work with a specific population of interest (C2, C11, C12, C17). Participants were also interested in gaining funding or learning how to write grants, which led to them joining grants even if the research was not directly relevant to their primary research agenda (C1, C2, C3, C7, C17). This was particularly salient for new faculty: \textit{"I was a brand new assistant professor. So it's just like, Yay, somebody wants to write me into a grant, I was... blindly ... saying yes to most things....  I was just excited to be invited... So my goal was just to like be on another grant proposal and learn about... grant writing because that was a really well-written grant"} (C2).

% \textchange{\textbf{Team level, research domain factors}}

\subsection{Conceptualizing research: factors that impact integrating goals of HCI and health researchers}
During the \textit{conceptualization phase}, researchers develop novel research questions, hypotheses, frameworks, and a research design that integrates across the disciplinary perspectives of the different collaborators \cite{hall2012four}. Most participants described their intended primary research contributions to be in the HCI field. However, this came in conflict with the contributions that health researchers expected to make. We highlight how some researchers were successful at conceptualizing research together, while others encountered barriers in having a shared integrated problem conceptualization. When such factors created barriers in the collaboration, it could lead to researchers joining projects that would not fit with their professional and career goals and expectations.  

\subsubsection{Jointly conceptualizing research}
Most participants had engaged in successful collaborations with health researchers in which they shaped research together. In most cases, one of the researchers in the collaboration had a research direction they were already studying to which they invited the other person, therefore the collaborator came in when the project direction was somewhat defined. 

However, few engaged in a symbiotic approach to defining the problem they would study together. Those who did, engaged in activities that help facilitate discussions around research directions, without imposing a particular research question. Some of the participants who closely integrated research processes were longitudinal collaborators who did several projects together over long periods of time (C2, C12). When they wanted to explore a new research direction some participants would conduct some research separately from their collaborator which they would then present and get feedback on: \textit{"I have like a nugget of an idea that I want to kind of explore... I want to just go do a few like, participatory design sessions, or... a few interviews just to... put these ideas in... designs in front of... actual clinicians"} (C2). In such cases, C2's collaborator provides feedback on a particular problem, which could later lead to a joint project. Similarly, C12 would present intermediate findings or ideas from a project that is outside of the joint collaboration, to assess the possibility of further defining a project jointly. C3 and C12 jointly defined a project with their collaborator because they both had been researching a similar problem from separate but similar perspectives, which made it easier to shape a project together.

\subsubsection{Limited Health researcher understanding of HCI roles}
When health researchers did not understand the role of the HCI researchers and of their research, they invited them to collaborate on projects after the research was already conceptualized. Participants often referred to such invitations as contributing in a "consultant" role to the project, in a way that was not directly in their area of research or did not have a clear HCI contribution (C2, C3, C8, C10, C14, C16). Some researchers were perceived as human factors or User Experience (UX) researchers (C2, C12, C14): \textit{"we just need somebody to redesign our EHR better, right? Human Factors person, like if you do that... wow, I don't do that"} (C12). Another way in which participants were perceived was as a programmer (C3, C7, C8): \textit{"you can program, this thing that I've envisioned in my head. And so that's going to be the scope of our collaboration is kind of me telling you what I know needs to be created, and you'll create it for us"} (C8 ). Other times the design was perceived as "\textit{artistic}" (C8). In such situations, the HCI researcher did not have room to shape the larger research problem because they were already assigned a problem that would not necessarily constitute an HCI research contribution.

HCI researchers were also invited to join projects that were misaligned with the work they wanted to do. This often occurred when participants were invited to contribute to later-stage research projects, rather than early on to conceptualize the project (C2, C14, C16). For example: \textit{"we're going to develop this whole thing, and we're going to bring you in year five, and you're going to do some testing, and then you're going to tell us [if] it's usable... No, that's not what I do. That's not how it works"} (C2).

\subsubsection{Limited understanding of the future of the research project}
HCI researchers sometimes engaged in projects for which they were not sure if there would be an HCI contribution. They knew at the time of joining the research they were not immediately going to have an HCI outcome, but hoped that after some time they would be shaping the research questions in a way that resulted in HCI contributions. This sometimes resulted in time frames of a few years without achieving a project scope that aligned with HCI contributions. 

Despite this lack of HCI contribution, participants chose to pursue these collaborations because it helped them form a relationship where they could do HCI work later: \textit{"establish the relationship... he's gonna be like...  grateful that we made him in the app, and then he's going to be interested in doing more interesting stuff with it... the app itself... [isn't] interesting. But you can imagine [ideas] I could probably do something interesting" }(C5). Despite C5 having been involved in this project for four years, they felt reluctant to keep going because the effort might not feel worth it: \textit{"it's been such a slog to even get through this basic app and to get him to do anything with it frankly, I'm not particularly... motivated to continue...  with the project because I think that it would be another two years of free tech development from my side. And... project is not funded"} (C5).  Other participants also had to navigate working on projects where they \textit{"gambled"} (C3) that there might be an HCI contribution (C3, C16), but later they realized there would be none. C3 took a year to realize that, and was now debating how to exit the collaboration: \textit{"It's not a failure... but it's not a project that's likely to lead to me getting a lot of my research questions answered"}. 

% \textchange{\textbf{Funding agency factors}}

\subsubsection{Funding agency expectations for roles and outcomes}
% todo say that the agencies impacted when people were involved in the research? or what was considered a contribution?
Participants formed collaborations with others based on the grants that funded their work, which impacted at what stage of the research the collaborator was engaged. The different funding models in NSF and NIH impacted how participants could engage with their collaborators and whether they could make their desired research contributions. 

The scope of the NSF and NIH grants created different motivations and levels of participation from both parties. C5 felt like the scope of NSF grants would not be of interest to their health collaborators because the outcomes were not compatible with expected health outcomes, which would make it hard to motivate the collaboration to them: \textit{"there's a big gap between what NSF funds and what NIH is and somebody who works in the NIH space, I think, what has trouble looking at this and seeing this as valuable... like, you know, running the study with 40 participants, like it's not an actual clinical trial so to them... does that evidence... really mean something?" }(C5). The scope of NIH grants meant that the HCI researcher's work was not always accounted for as part of the grant (C2, C16). C2 received advice to not add HCI work in the grant because it would be received negatively: \textit{“I kept wanting to add more [of their own research] into that aim... you don't need to do that's... probably... gonna mess things up with an NIH [grant]... you really are just the app person. And that’s all you need to say in the grant" }(C2). 

% this feels a bit repetitive
Participants found it challenging to innovate in technology when they felt that was not the right solution for the problem: \textit{"HCI community, the NSF community values innovation. However, simple text messaging interventions work just fine. They've been shown very... effective with many populations... They're not innovative. So I can't really...  publish it as a CHI paper"} (C12). C1 also found that funding agencies in  Health expected the  use of established technologies, rather than novel and risky ones: \textit{"NIH ... they actually prefer not to involve brand new technology. Because it's too exploratory. It's too much unvalidated... So they're more they want to see more confirmatory results... That's novelty, whereas for me, confirming is not very interesting. I want to see... what's new, and what hasn't been confirmed" } (C1). 

% \textchange{\textbf{Individual factors}}

\subsubsection{Researcher's seniority in the research community}
Although mentors facilitated many successful collaborations for our participants, mentors also impacted how the HCI researcher engaged in the collaboration, particularly in cases when the collaboration did not progress as expected. C11 thought that, because of their junior faculty status, they had to compromise on their research ideas to maintain a good relationship with both their health collaborator and their senior mentor who facilitated the two meetings.  C11 felt that having received tenure would allow them to have more autonomy in defining projects with their collaborators: \textit{“I feel like I can speak more, and I can be, I can have a more active voice in what I'm interested in”} (C11).

\subsection{Implementing research: using HCI methods in collaborations with Health researchers}

During the \textit{implementation phase} research teams launch and conduct transdisciplinary research \cite{hall2012four}. During this stage, researchers collaborate to refine and extend research questions and methods used, and execute the research. However, in practice this process is difficult and not necessarily successful. In this section we describe the tensions in integrating methods, approaches and timelines, encountered by HCI researchers when working with health researchers. Several participants (C5, C9, C10) described how the long term goals of the collaboration were aligned, but tensions arose in their misaligned approaches: \textit{“the goals were the same...but I think some of the orientation towards that and some of the expectations of how we get there were misaligned. And for me, that did create a lot of tension”} (C10). 

Participants encountered resistance from their health collaborators with regard to the methods they used, which reflected the difficulties of using a human-centric approach in their work, motivating the rigor of methods used, justifying that the work done is research, and aligning practices around how methods are implemented.

\subsubsection{Timelines for executing research}
Participants described health projects driven by their collaborators, often funded through NIH type of funding, as having much longer timelines than HCI projects are typically expected to have, which can go as long as five years (C3, C7, C11, C12, C15, C17): \textit{“In the HCI community, if you publish papers every five years, you're not going to make a lot of progress as a researcher} (C12). Because of that, the HCI participant's involvement and outputs from the work were constrained by the timeline of the grant. Participants described the timeline of five-year projects as having variable engagement: \textit{"I'm seeking to have... an equal role... if we look at it over the five-year collaboration... even though there were periods where I was... getting more of my research questions answered, overall, it probably works out to where we were both getting our work... done"} (C3). The way in which this was reflected in C3's work was through sometimes favoring HCI research outputs and papers, and other times acting more as a consultant or HCI practitioner:\textit{ "at the moment... they're just building out the tool... what I've done in the last three months... a cognitive walkthrough of the prototype... that project... may morph into... next big thing is... we have evidence now that we need to actually do some more intense HCI research" }(C3). 

The long timelines of health projects were also challenging because focusing only on a project with health outcomes would mean too long of a wait for publications. HCI researchers, therefore, had to be strategic about their publications about these longer-term projects: \textit{“In the HCI community, if you publish papers every five years, you're not going to make a lot of progress as a researcher… I wanted to continue publishing... I had my expectations, how to put together a tenure portfolio. I had to be very creative about what parts of longer projects can be published sooner before the projects are done”} (C12).

% … I wanted to continue publishing and goals, but I had my expectations, how to put together a tenure portfolio. I had to be very creative about what parts of longer projects can be published sooner before the projects are done.”} (C12)

\subsubsection{Understanding of HCI methods and rigor}

HCI researchers experienced a misalignment in research approaches by having some of their methods (e.g. qualitative research, design methods) not taken seriously, not considered to be research, or not considered to be rigorous (C2, C8, C9, C10, C14, C15). This made it more difficult to conduct their research. 

\paragraph{Skepticism of qualitative research}
Participants shared the challenge of health researchers considering qualitative work as less important than quantitative work: \textit{“early-stage work, my clinical collaborators don't even consider that research...because they're interested in interventions and in technology, it means they're naturally interested in the outcomes of that. And so to them, that's the data. That's the research”} (C2). C8 found that collaborators were more receptive to engaging with a user-centric approach when it has a quantitative component: \textit{“that reluctance to engage with people through sometimes more qualitative methods... Some people definitely have a bias towards, you know, if it's a survey and it's numbers...they'll be receptive to us maybe doing some sort of engagement on that level. But if it's more qualitative, or feels anecdotal...sometimes there can be some hesitancy”}. 

\paragraph{Different views on what is considered data.}
The resistance to qualitative methods and design was also reflected on what was considered data or not. C17 found that collaborators thought that qualitative data did not consist of actual data:\textit{ “This is just qualitative data. Aren't these just anecdotes anyway? And no, they're not anecdotes, right. This is real data”} (C17). In response to collaborator' skepticism about the rigor of qualitative design research,  participants worked but sometimes struggled, to establish what counts as data early on (C2, C15, C16). For example, C16’s collaborators were in agreement about conducting formative research, but did not include this type of research in the IRB application: \textit{“I'm snapping pictures, I'm gathering all this visual content to try to inform how we're going to design this up and talking to people... going through and doing some sort of analysis on the images...and the research coordinators [said] you can't do that...I just had this sinking feeling of like, I can't use anything that I've gathered, like, I can't do anything with any of this. And we ended up having this conversation about what counts as data, and what was actually approved by our IRB for us to collect."} (C16). C16 attributed this to the focus of their collaborators on quantitative data. 
% in the “it never occurred to me that you would write an IRB that involved any sort of fieldwork.. that you wouldn't be a you know, your data collection wouldn't start this second, you got off the plane.. but as we talked, I'm like, Oh, yeah, this is like this is ethnography. Right? This is participant observation. The way that we write IRB is for that. They wrote this IRB as a trial, right? To them the data, or the quantifiable responses to a structured set of questions about do you like this? Or do you like this?” (C16)

% \paragraph{Other criteria of assessing research rigor.}
% HCI sample sizes were considered too small: \textit{“in medical domains, they tend to do things in a much more systematic way [and] need to apply a level of rigor that is much more...we have a small sample size, but [collaborator's grant proposals] have hundreds if not 1000s of participants”} (C9). C14 also pointed out that the publication venue of a researcher signals their rigor, which can be hard to interpret given conference publication in HCI: \textit{“if they look at [an HCI] CV, and they're like, hey, a bunch of conference proceedings... they're not rigorous”} (C14).

\paragraph{Design methods not taken seriously}
Participants also encountered resistance to design methods: \textit{“sometimes when they don't take design methods seriously, unfortunately… they think a lot of the design stuff is just… a lot of mumbo jumbo kind of being creative”} (C8). C2 thought their collaborators did not take drawing activities seriously: \textit{“they don’t get why we’re all drawing pictures…. you feel like it's a feel infantilized... And that was very reminiscent to me of how  these clinicians were looking at me”} (C2).

\paragraph{Centering user needs in tension with clinical expertise}
Participants mentioned that one of the barriers in doing research with a health collaborator is to integrate the perspective of users in the process of designing a technology-based health intervention (C1, C2, C10, C14). In some collaborations that participants engaged in, health researchers approached problems by already having a solution in mind. This made it difficult for the HCI researcher to follow their research process of designing a solution that was based on formative research or having to compromise on user-centric features (e.g. types of data the user wanted to track, but the clinician did not see value in). 

C10 thought that a lot of the tensions between balancing the user’s perspective and clinical perspectives came from the training of the health researchers: \textit{“they're trained to be the ones in the position of power...[user-centered design] was kind of antithetical to their training...it really was, again, kind of like an entire, like, shift in their worldview, and I think that's why there was a lot of pushback.”} C10 also attributed these challenges to ableism: \textit{“ableism is, it's not the app that's the problem. It's ableism in the entire system, that's the problem”}, which leads to not recognizing the patient perspectives. 
% C1 found that clinicians might not be interested in incorporating user needs because of their focus on outcomes, rather than on design of technology: \textit{“clinicians don't care about how things are being done. I mean to kind of overgeneralize to simplify things. They, at least in my context, I think they care about like specific outcomes.”} (C1)

% Consequences of not including health researcher data
When the health expertise took first priority, participants thought it came at a cost to designing tools aligned with the needs of the intended users: \textit{“they [collaborator] weren't choosing technologies that were aligned with our communities”} (C14). At times this meant not designing features because they were not aligned with a way of representing health constructs: \textit{“the [health researcher] came back with like, [feature] isn't even a clinically valid measure. That's useless, like [feature] doesn't mean anything [clinically]… so that shouldn't be in the app”} (C2). Because of this, the technology might have not included a functionality that users found important to their needs.  

Because of the difference in clinical expertise between the health and the HCI researcher, that can lead to a reliance on the health researcher as a representative of patients or of other health practitioners’ perspective and can take the technology into a different direction than that addressing patient’s needs (C2, C4, C7). This can be helpful in some situations, where the health researcher becomes the \textit{“liaison between us and then the patients”} (C7). On other occasions, the over-reliance on the health researcher can lead to decisions that do not match those of users more broadly. C4 relied on the health collaborator to understand the problem space, which led to a design that did not represent the user’s needs: \textit{“what we started with was influenced by what's that one [health collaborator] told us… it was eye-opening….then talking with other [participants] and realizing they're doing it very differently.... we spent a lot of time working with that one person... if we have started with the interviews... it would have been a slightly different [tool]”} (C4). 

% For several participants, their collaborators were either not interested in incorporating feedback from users in their research, or it was not part of the project grant, and therefore was not a priority (C2, C7, C14). The lack of interest in incorporating user needs came in the way of C14’s ability to do formative work. C14 found that the focus of their collaborators was on wanting to translate a health intervention into a  digital format without regard to participant needs: \textit{“we just want to translate this in a digital format... from my perspective... I want to go back and... think about  the needs, and they're already... a few steps... ahead in terms of just wanting to translate whatever they have”} (C14). 
% C7’s collaborator did not understand why they needed to speak to users at all, and not before the evaluation stage of the intervention:\textit{ “he really didn't understand why we need to talk to patients.. before the evaluation or as part of evaluation, because they were... from the [field name], where everything is about much metrics, and more quantitative oriented research.”} (C7)

\paragraph{Artifacts facilitating knowledge integration}
Participants explained how the collaboration required a lot of building common ground. To align perspectives about methods and approaches, participants used different approaches to introducing their collaborators to methods through hands-on activities practicing methods, showing examples or demonstrations of methods or technologies. 

% Participants mentioned that building common ground about certain methodological approaches took months at times (C10, C11, C14): \textit{ “the first.. six to eight months was just like a lot of onboarding in terms of understanding kind of what the goals of each party were and understanding how to even speak to one another” }(C10). 

Some participants asked the collaborators to experience the method or approach. C10 had the collaborators participate in design activities to learn about them and to understand the rigor of the methods:  \textit{“There were a lot of challenges in getting them to loosen up...we're really just gonna have them sketch things out... get them to... understand a different kind of rigor in the research”} (C10). C10 found it useful to describe the entire research process through repeated presentations and examples: \textit{“once they got to the end product of...this is what a CHI paper looks like...all the steps we've taken up to that point became more clear to them...'Oh, I see why we were like collecting all these sketches' or...now let's see where they go and what they are for.}”(C10).

Learning by experiencing also helped collaborators understand the capabilities of technology. For example, some collaborators expected technology to have much more advanced algorithmic capabilities. C12 mentioned that when they had a working prototype for their project, C12 asked the collaborators to use the technology: \textit{“I first offer to them is here, try this out....and just to get a feel for what it's like.”}. This helps set the expectation of technology capabilities and what C12 can contribute through technology.  However, C12 discussed challenges with using prototypes because they are not always readily available for new problems, and having a suitable prototype for a particular domain requires formative work.
%  : “But here's where it gets tricky to develop a mock up, I kind of need to understand something about what their needs are. Because if our mock up doesn't match what they need, that that will be not a good experience for anybody. So that's why we came in with vague descriptions. But I think we were really lacking something tangible that they could react to and say, Ah, this is what, how it will be.” (C12) 

To explain why a method is useful, some participants used examples from their previous projects, for example: \textit{“here's either a past project from, say, our portfolio, where if we hadn't done things, the way we're championing doing them right now, we would have arrived at a really not only wrong solution but could have...really done more harm than good.”} C8 also shared that\textit{ “people pulling in ... [examples from] popular media and stuff like that...can make it a little more tangible”} (C8). 

% Most participants thought it is part of their role to educate their collaborator about the methods used (C1, C3, C7, C8, C10, C12, C14, C17).   C3 saw this as the health researcher being a client: \textit{“it is surprising to me to reflect on how similar sometimes your generic clinical collaborator feels to a client some ways...in terms of trying to educate them about the value of doing these sort of small and qualitative, formative pieces of research upfront”} (C3). C10 found it challenging to explain what they considered basic methodology to senior researchers: \textit{”like all the stuff you get trained to do as a grad student...it was weird having like seasoned researchers that needed to be kind of retrained in some of that”} (C10).

\subsubsection{Differences in the terminology of methods}
HCI researchers were thoughtful about what terminology to use. Researchers avoided  common HCI terms because they represented approaches that were not well regarded by the potential collaborators, or had a different meaning: \textit{“I think initial challenges were different language barriers because they have a lot of jargon, we have a lot of jargon and so it was like a lot of figuring out what even the other person was saying”} (C10). Some of the terms that created tensions were \textit{exploratory}, \textit{user testing}, or \textit{implementation}.

\paragraph{Approaches to sharing terminology}
To translate approaches used in HCI, some researchers were very careful about terminology they would use. C1 avoided using the term \textit{”exploratory”}, as advised by one of their mentors because it would not be perceived as valid research: \textit{“you should never say exploratory, because, you know, in this world, they don't really appreciate exploratory”} (C1). Instead, C1 would look for other terms that would fit some valid terms that health researchers might be familiar with:\textit{ “there's actually plenty of research being done in the clinical world that has the same goal that I have. And so there's formative research, that's actually legit, and that in, like, clinical world, and it's called… community-based participatory research… there's all these methods that they think is valid, that are essentially like participatory design”} (C1). 

The meaning of terms related to evaluation was also challenging to use. C1 would replace \textit{“user testing” } with terms that were more familiar to collaborators: \textit{“I use a different word, but they're very used to saying, usability testing, or human factors testing, and the sounds... validated. And so I... use those terminologies to make them understand what we're doing”} (C1). C8 encountered challenges in using the term \textit{“efficacy”}:  \textit{“What counts as efficacy comes up quite a bit...for HCI people efficacy means almost kind of like the proof of concept, or a measure of effectiveness. And that could mean so many things...But when maybe you're talking about health science colleague... efficacy,  has very specific requirements about... thresholds that need to be met according to this very specific measures.”} (C8). C16 had developed new terms to speak about evaluation to their colleagues: \textit{“I don't say we're going to evaluate the prototype... I say we're going to see what happens when that prototype gets appropriated into a real-world setting”}. C16 explained their reasoning because \textit{“It's really important to separate out the evaluation of the interface and the information design and the information architecture from the impacts of the clinical intervention.”}.

% Another challenging term to use was \textit{“implementation” }because it could have different meaning to HCI and to Health researchers: \textit{“when my clinical partners talk about implementation, they're thinking more what I might call deployment... that human element of actually adopting and using some of these things we've been engineered”}. (C8)

\subsubsection{Differences in achieving desired research outcomes}
 
Because of the misalignment of approaches and methods between HCI and Health collaborators, HCI researchers adapted their approaches. They resorted to conducting their research on the collaborative project outside of the scope of the collaboration (on their own) or had to adapt methods and approaches to fit the collaboration needs. 

\paragraph{Conducting HCI research within but outside of the collaboration boundaries}
Because some participants encountered barriers in having their formative work or design work recognized by collaborators as part of the scope of research, they needed to conduct formative or design research outside of the scope of the expectations of the jointly defined project or grant (C2, C7, C16). For example, C7's collaborator did not see the need to conduct design research. Thus C7 pursued this research on their own, without support from collaborators. C7 found it challenging to not have the collaborator’s support in recruitment: because the user-centered research was not included in the grant, it was also not part of the IRB, and therefore the collaborator did not engage in recruitment, leaving C7 to find other channels of recruitment for patients. Although this research was done independently of the grant's resources, the publications resulting from the work were included in the annual reports of the grant. This tension was accepted by C7, because of the desire to be able to engage with patients later on: \textit{“what can you do? Then you can't work with patients”} (C7). After this experience, C7 decided not to pursue future collaborations with this researcher after the ongoing multi-year project was complete.  

C2 also needed to work on papers outside of the collaboration scope, because their collaborators did not initially consider the research to be publishable:  \textit{“formative stuff that I do… I write up... send them those drafts...  they're like... this is interesting. I didn't even realize you could write a paper on this. And then I go and get it published”} (C2).

\subsection{Translating research: factors that impact translating research to the real-world}

During the \textit{translation phase}, researchers create pathways to apply their findings to real-world problems \cite{hall2012four}. This creates tensions because the research outcomes of HCI and health researchers are misaligned in this perspective. 

\subsubsection{Tensions between innovation and real-world impact.} C12 also described the difference between the disciplines also as driven by a desire to solve concrete current problems in Health, versus solving pushing innovation forward in technology: \textit{"technical innovation ... I really enjoy doing this kind of work. But it is somewhat disconnected from solving concrete today's problems... It's really hard to do both to be a researcher... who tries to do transformative technological research, which is what NSF expects and at the same time be somebody who kind of makes that translation of this work into practice"}. 

Health clinical trials were not seen as advancing the technology innovation agenda and contributing to the HCI field, but some participants felt it was important to engage in such research because it had the opportunity to make an applied impact (C1, C7, C9, C10, C12): \textit{"actually improving... health outcomes. It takes years of work. And so I think there's like that sort of sense of accomplishment when I can actually have my work impact, improving outcomes." }(C1). C1 saw being disconnected from the broader HCI community as a consequence of this work:  \textit{"I feel a little outdated in terms of what's new... in the field of like CHI and CSCW... I feel like... I'm lagging a little behind, because I'm focused on like, work on the ground... outcomes-oriented research, but I'm okay with that" (C1). C1 felt they could engage in this work because they felt like they still did "quite a bit of novel things"}  through their HCI work.

\subsubsection{Training in translating research to practice.} C12 explained that \textit{"HCI researchers are ill-prepared to handle [clinical trial] because we don't really learn much about clinical trials as part of our PhDs in HCI"}, which would make it difficult to do the extra steps of translating the innovation work to real-world problems. C12 also emphasized that the translation work to implement a solution can involve a lot more than just clinical trials: \textit{"it requires more collaborations, rather than trying to be the... one-man show and the person who does everything... this is outside of collaborating with health scientists. This is collaborating with like IT departments of hospitals, and groups that can actually... help to implement solutions once they've been tested and shown promising."}

%% file: 05discussion.tex
\section{Discussion}

% - team science encourages us to work on integrating research across field for maximum innovation
% - we should examine our practices; HCI is interdisciplinary but if we wanted hci researchers to be successfull in transdisciplinary initiatives, there's a need to do more to support them in conducting this reasearch

In this paper, we illustrate the challenges that early- and mid-career faculty in HCI experience in integrating HCI perspectives in collaborative work with health researchers. These integration and knowledge challenges have been studied in Team Science and in current reports and models of cross-disciplinary research \cite{national2015enhancing}. Past research in Team Science offers effective recommendations on how to better integrate perspectives across different fields. The results from our study reveal the need to examine HCI research practices and expectations in cross-disciplinary collaborations. Based on our findings, we discuss how HCI researchers and institutions can incorporate recommendations from Team Science. 

Our study adds a new lens to investigating cross-disciplinary collaborations. CSCW researchers have previously investigated how scientific teams interact as well as their processes, use of language, and use of technology to integrate knowledge \cite{olson2000distance,olson2013working,mao2019data}. Such research can be effectively used for supporting research teams in developing collaboration, coordination, or role division for effective team functioning. However, a Team Science lens enables researchers in HCI to examine how research is integrated between different fields and what influences the success of integrating perspectives across different fields. Hall's model of transdisciplinarity allows us to identify participation, representation and integration of research perspectives at different stages in the research process \cite{hall2012four}. Considering factors that impact collaboration within the context of individual teams and academic institutions can offer HCI researchers a framework to better integrate HCI and health research in the varying academic environments and circumstances where they create technology. 

Our results indicate that research integration across HCI and Health can be difficult to achieve throughout the different stages of assembling a team, conceptualizing research, implementing it and translating it into practice. In this research, we use a lens of \textit{transdisciplinarity} to assess collaboration. The intersection of Health and HCI presents a complex problem space in which team science and transdisciplinary research have the potential of producing innovation \cite{spring2012emerging}. However, our findings point to situations when researchers might not find tight research integration desirable. For example, HCI research might not be desirable for some health researchers when they need to follow tight grant deadlines, or when they need a design consultant or programmer. On the other hand, some HCI researchers might not find it desirable to engage in the translational stage of health research. Additionally, collaborations through which a research member a team member is brought in late in the research process can miss out on joint conceptualization of the problem space and inclusion of cross-disciplinary research approaches. Our participants showed an interest in achieving research integration and participation from both HCI and Health, even though sometimes it came with a cost, such as lengthy upfront investment in a collaboration with uncertain outcomes. We draw on the participant intent in building research integration with collaborators in our discussion. Below we use a neutral terminology of \textit{cross-disciplinary collaborations} to present considerations that can support teams in more strongly integrating research perspectives between Health and HCI researchers.

\subsection{Team level support for research integration across HCI and Health}
Successful collaborations can be developed through learning collaboration practices across fields and through tools that help individual members of teams to better communicate with each other. We highlight how aligning perspectives can be achieved through training, building methodological common ground, and supporting communication.

\subsubsection{Supporting cross-disciplinary training}
Our participants expressed a need to get training support in navigating research partnerships with health collaborators. Despite completing PhDs in which most participants worked closely with health researchers, doing so was unfamiliar for early-career researchers while being in a lead investigator role. Team science initiatives propose the importance of training programs in which scholars develop competencies in navigating knowledge from other fields, navigating relationships, and coordinating teams \cite{national2015enhancing}. Training grants already exist for health researchers through the NIH \cite{nih-training-center}. However, our participants noted that training grants were inaccessible or incompatible with the promotion expectations in tenure track positions, coming in conflict with teaching expectations. This indicates a need for the HCI field to design training programs that support researcher training, throughout the PhD and after, to be fully equipped to enter a professional space where they lead cross-disciplinary projects. Our participants identified the following areas as being valuable for training: (1) how to translate HCI methods to health researchers (e.g. design methods, qualitative research, data types); (2) Health research methodologies, such as how to conduct randomized controlled trials; (3) appropriate terminology and how to use it with health collaborators (e.g., communicating exploratory research, evaluation criteria, the meaning of user testing); and (4) an overall understanding of the role that a health collaborator plays throughout the research life-cycle and strategies for how to have a collaboration that is meaningful for both parties. Such training can help HCI researchers in starting collaborations, as well as maintaining them. Further research is needed to identify a range of competencies that new investigators can benefit from. Such a list might be further expanded with the support of scholars with more experience who developed strategies to navigate cross-disciplinary collaborations over time.

\subsubsection{Building methodological common ground through translational resources }

One of the needs that emerged in our study was to better translate HCI methods across disciplines. How to translate these methods will vary based on the collaborative domain. In the context of HCI-Health collaborations,  HCI researchers needed to justify their methodological rigor and approach to health researchers. Past research shows that boundary artifacts could help teams in developing shared understanding \cite{salazar2012facilitating}. Team members can leverage successful strategies and artifacts that were identified by our participants: developing and sharing examples of how HCI approaches improved health outcomes and enabled more effective solutions; demonstrating how certain design activities (e.g. storyboarding) are beneficial for later research steps; or sharing technology prototypes that health researchers can use to make the potential of technical innovation more concrete for collaborators. Additionally, researchers have proposed the use of translational resources that support practitioners in applying a new knowledge domain to their work (e.g. using theory for design practitioners, or rethinking assumptions about personal informatics) \cite{kirchner2021they,colusso2017translational}. Such resources could be developed through design cards, toolkits and materials that are incorporated into the research process \cite{colusso2017translational,colusso2018behavior,colusso2019translational,nadal2022tac}. 

Another valuable translational resource is scholarship that highlights how methods can be adapted or translated when conducting research to design tools at the intersection of Health and HCI. For example, researchers have developed guidelines for publishing qualitative research in Health Informatics \cite{ancker2021guidance}. Building shared vocabularies can also help researchers to establish common ground \cite{olson2013working}. Researchers have started building such shared vocabularies between design and implementation science methods \cite{dopp2019glossary}. Participants in our study highlighted terms that needed translation in their collaborations, such as \textit{exploratory, user testing, evaluation, implementation}. However, more research is needed to develop essential vocabularies that could help researchers more easily communicate about research. 

Such translational strategies serve as a model for the HCI and health community. The development of similar resources could aid early-career researchers in having conversations with health researchers about the use of design methods and participation in projects (e.g., jointly conceptualizing a problem space). 

\subsubsection{Supporting communication about goals and outcomes across disciplines.}
HCI researchers struggled at times to know if a team or project will be a worthwhile collaboration. Existing Team Science toolkits can be leveraged to align perspectives and guide discussions about shared goals or outcomes \cite{spring2019continuing,teamsciencetoolkit}. These toolkits include prompts for conversations about the goals and outcomes of the collaboration, roles, and contributions of the participants, as well as deciding which team members will make what decisions and how \cite{teamsciencetoolkit}. Researchers in HCI have developed similar types of tools for supporting remote team collaboration \cite{bietz2012improving}.  Such toolkits could also help team members better identify the knowledge, skills, and attitudes of the team members early on in the research lifecycle \cite{hall2012four}. For example, HCI researchers can facilitate conversations with health collaborators about their career goals and whether ongoing projects might be adjusted to suit goals of both parties. To support alignment of expectations in working together, researchers must discuss their goals (e.g., scientific, timeline), division of responsibilities (e.g., individual/team contributions to doing research, writing responsibilities for reports, team membership, and decision making), authorship and credit (e.g., what manuscripts will be written, how will credit be given and decided, who will deliver presentations and answer inquiries), communication approaches (e.g., how will communication be conducted, how the research agenda might get redirected, how will new collaborations and spin-off projects get negotiated), and conflict management plans (e.g., encouraging diverse perspectives, managing conflict constructively) \cite{teamsciencetoolkit,hall2019comprehensive}. Such conversations might be adapted to specific situations that are specific to conducting HCI research, which might involve expectations about: what approaches will be used in the research (e.g. inclusion of a user or patient centric approach), what data to collect and when for publication, starting with the IRB; timelines to HCI publications might be shorter than in Health, might involve different type of data and might be published in different types of venues (e.g. conferences vs journals); research responsibilities (e.g. development of technology and corresponding timelines); responsibilities that an HCI researcher might have outside of research that differ from the collaborator (e.g. teaching). Early-career researchers might have promotion expectations related to the above that are different than their more senior colleagues that can be useful to communicate. 

\subsection{Institutional support for research integration across HCI and Health}
At the institutional level, the Team Science field recommends creating initiatives and incentives for investigators to find collaborators, normalize transdisciplinary research, and recognize the effort of participating in collaborations for promotions and tenure \cite{national2015enhancing}. 

\subsubsection{Supporting finding collaborators}
Our findings show that early-career researchers are particularly susceptible to difficulties in starting new collaborations, dependent on the type of institutions, mentorship, or institutional structures that promote cross-disciplinary collaborations. The HCI community can support early-career researchers who might be struggling to establish collaborations in academic environments with limited resources. Physical proximity, for example, is a key opportunity that HCI communities are well-positioned to encourage via conferences \cite{binz2015making}. For example, symposiums that alternate between HCI and Health Informatics conferences bring together researchers from HCI and Health \cite{weibel2019symposium}. Establishing more formal connections with other conferences in the Health field, organizing regular joint events that promote meeting health researchers, and including skill development events that help HCI researchers communicate the value of their work across disciplines using analogies, metaphors, or lay language instead of discipline specific jargon \cite{vogel2012influence}, are all ways to build upon these initiatives. 

Brokers -- well-connected individuals in the community \cite{hall2018science} -- can play a particularly important role in leveraging the strength of their networks to facilitate cross-disciplinary connections and mentor early-career faculty, especially individuals who identify as part of marginalized or underrepresented groups. For example, previous initiatives for women young faculty members show that the majority of the scholars getting mentored successfully acquired competitive grants \cite{nagel2013building}. This measurable success shows the value of developing opportunities for specific research groups in the HCI community to establish meaningful professional relationships with the brokers of the community. In addition, the HCI community can adopt Team Science approaches such as online networks of researchers to facilitate finding expertise \cite{kahlon2014use} or training programs that connect researchers \cite{mhti}. These opportunities could ensure early-career faculty build networks, leverage proximity, and explore potential collaborations to advance their research agendas.

Institutions must remove obstacles to doing research across disciplines \cite{national2015enhancing}. These obstacles can be removed by creating funds for transdisciplinary seed grants, which our participants valued for building collaborations, and intentionally merging departments to foster the intermingling of faculty \cite{national2015enhancing}. Institutions can also support researchers in establishing  research centers that foster cross-disciplinary research and train students \cite{national2015enhancing}. Several NIH-funded transdisciplinary centers already exist \cite{nih-hci,nih-training-center}, but increasing the support for such centers within the HCI field can increase the capacity of developing researchers and research that tightly integrates and innovates at the intersection of HCI and Health.

\subsubsection{Attributing value to cross-disciplinarity collaborations through promotion policies}
HCI researchers have raised questions about how to balance HCI and Health contributions toward promotion. HCI contributions, in contrast to health contributions, have been discussed in prior work. For example, researchers have debated the role that HCI researchers should play in the evaluation of systems such as randomized control trials \cite{mamykina2021scaling,klasnja2011evaluate}, but no clear consensus exists on how that is valued for researcher promotion and academic career advancement.

To encourage participation in cross-disciplinary research, Team Science recommends establishing explicit language about this research in expectations for promotions and tenure. Although the status of such initiatives in the United States is still limited \cite{hall2013recognition}, researchers have outlined policies that institutions can follow to recognize team-based work \cite{national2014convergence}, such as  allocating individual credit for teamwork, diversifying the attribution of contributions, placing value on research outcomes, making creative work available to others and the public, and minimizing disputes over project ownership. Such norms are adopted in some health departments \cite{teamsciencetoolkit-uw-promotion}. Computing researchers have been recognizing the importance of collaborative work in promotion \cite{shneiderman2016teamwork}. Considering the HCI community as an institution, members of this community can develop language for academic departments to recognize the different types of contributions, and often invisible labor, of HCI researchers. Transparent discussions and agreements must be reached about what contributions each research community values in addition to revising these values as they evolve over time. 

\subsubsection{ Aligning funding expectations with methodological integration}

While funding agencies such as the National Institutes of Health and the National Science Foundation are increasingly providing support to research that involves both Health and HCI \cite{nih-hci,nsf-sch}, these institutions still need to to support the diverse research of both communities. Because funding agencies heavily influence research incentives, and grants increasingly require researchers to have HCI or health collaborators, it is important for funding agencies to support the goals of both types of collaborators as part of the grant.  

The tensions encountered by our participants suggest a need for NIH grants to better account for the type of contributions that are meaningful for the HCI field. In cases where National Institute of Health projects require the expertise of UX practitioners, expectations about what role an HCI researcher has on the project as a co-researcher, rather than a consultant, must be transparent and inform team decision making. To achieve transdisciplinary research, funding agencies should  support the development of innovative pilot projects, ideas that emerge during collaborative projects \cite{hall2012four}, or real-time adjustment to projects as they unfold \cite{national2015enhancing}. Expanding the focus of transdisciplinary research funding could better enable formative and impactful HCI research.

\subsection{Reflexivity about research integration across the research lifecycle}
% some info should be here about why we will talk about reflexivity
Reflexivity has been growing in HCI as an effort to add transparency in taking responsibility and being held accountable for the research process, or to critically reflect on the research process \cite{liang2021embracing,malinverni2017autoethnographic}. The artifacts that researchers create are embedded in their sociotechnical context and are impacted by the politics of the infrastructure in which they are made \cite{winner1980artifacts}. Reflexivity encourages self-reflection and understanding of researcher's biases and role in the research and how that might impact their work \cite{liang2021embracing}. Our findings show how imbalances in collaboration and in integration of researcher perspectives can impact the output of the research through biases in how HCI or health researcher perspectives are included, the methods used, and how researchers consider the applicability of technology to the real world.

% epistemic perspectives 
\subsubsection{Reflexivity about whose views are integrated into research and when}
Hall's model can help HCI researchers be reflexive about the role they and their collaborators play in the research process. Researchers can use the different stages of research to reflect on how their epistemic perspectives are incorporated, the extent to which they participate in different research activities, or what methodologies they use. We identify several situations when a researcher's perspective and participation can impact the technologies being designed. For example, when HCI researchers are brought late into a project, it can be difficult for them to incorporate human-centric methods or to ensure the research problems and technology design meet people's needs. On the other hand, when HCI researchers do not integrate the health researcher's perspective early on, or do so in a superficial way (e.g. for access to study participants), the technology design might not suitably address an actual health problem. When both health and HCI researchers contribute perspectives in the design process, there is a chance to overly represent the health researcher's expertise over that of participants, because access to participants can be difficult. Our findings show that forming diverse teams can be difficult to achieve, which can make it difficult to mediate the imbalances in participation.

Given that integrating different disciplinary perspectives can impact the research and artifacts that teams create, we propose that research teams use a reflexive approach of reporting their research process. In line with efforts to increase research transparency, researchers could document participation, roles, and contributions of themselves and health collaborators in the research process at different stages of research. Such reflexivity could be added to method sections of papers or could be reflected in an author's contribution statements of their roles in the research \cite{maruvsic2011systematic}. Through such statements, researchers can be more transparent about how different perspectives were prioritized in the work, at what stages of research, and through what type of participation. Researchers could include decisions made in the research or design process that led to scoping the design space, making design decisions, including or excluding technology functionalities that affect the outcomes of the technologies created. Reflexivity statements can provide context in interpreting the research in light of the varied participation.
%maybe say this could impact how the research is interpreted, adding context to the research itself and what perspective impacted the outcome. 

% maybe being reflexive in the research process ?? !!!!!!

\subsubsection{Reflexivity about applicability of research to practice}
One tension we observed in integrating research across HCI and Health was the misaligned expectations in translating research into practice, including translating interventions in clinical settings, or creating technologies that can be deployed for multi-year trials. While this translation was a priority to some of the health collaborators of our participants, it conflicted with the disciplinary expectations of HCI researchers. Applying a technology to a practical setting did not necessarily result in research contributions that were rewarded in the HCI research community. In contrast, solutions that are considered innovative in HCI research are not always appropriate or applicable for the clinical context. To address this potential misalignment with health practices, HCI research papers could include a section that discusses the concrete applicability of the technology developed into real-world settings (e.g. clinical settings, applicability to different health conditions), its fit with the practices of the target setting and participants, or an assessment of considerations for potential future translation to practice.

% TODO we can add that it is helpful to visualize the research lifecycle and how methods that are non-traditional to health researchers (sketching) are beneficial later in the process; show examples of how considering users concretely improved health technologies

% A Team Science lens helped reveal opportunities to address the challenges of achieving transdisciplinary research at the intersection of HCI and Health. In this section, we examine recommendations from the Team Science discipline on how best to support transdisciplinary research. We draw from a recent national report reviewing best practices in supporting team science \cite{national2015enhancing}. We present and adapt these recommendations--at the team and institution levels--to the context of research in HCI and Health. 

%% file: 06limitations-conclusion.tex
\section{Limitations and Future Work}

Our study presents new considerations for how the HCI community can support early- and mid-career researchers in conducting innovative and impactful transdisciplinary research in health. However, we also acknowledge the limitations of the study and opportunities for future research. It is important to understand the perspectives of health researchers on their challenges and strategies for collaborating with HCI researchers. The larger aim of this work is to expand its scope and compare the perspectives of both HCI and health researchers to develop and evaluate tools that bridge the gaps between the practices of both types of researchers. However, more work is needed to develop both comprehensive research and official guidelines for individuals, teams and institutions on how to better support cross-disciplinary collaboration in HCI, such as the ones created by CRA \cite{boules2016future}.

The participants in our sample include HCI researchers who secured positions at academic institutions. Many participants were located at research-focused (e.g., R01) institutions in the United States. Therefore, participants' perspectives are not representative of early-career HCI researchers outside of the United States, researchers with experience in HCI and Health who hold non-tenure-track academic positions, or researchers in HCI and Health whose work is in a commercial or industry setting. Researchers in each of these roles are part of the HCI community, and future work must shed light on their perspectives and address their collaboration and career needs. \textadd{In addition, participants in this study were recruited primarily through the proceedings of the CHI conference, though not exclusively. While this is a common publication venue for early career HCI researchers in the US, there are certainly researchers conducting HCI research who might not have been included in our sample, which include researchers who publish in different venues, who did not publish recently in the venue, who work in departments or at institutions where CHI is not a primary publication venue, or who just transitioned to the HCI field. Overall, the participant sample has a bias towards top tier institutions, likely to be better resourced and that value specific publication and research outcomes. Future research investigating contexts outside the US might have to use a different recruitment approach to account for publication patterns and venues, types of institutions, and types of academic faculty roles that may vary in other countries.  }

\textit{Understand collaborations in different infrastructure settings.}
More research is needed to understand how the infrastructure of academic collaborations outside the US are shaped. While the methodological challenges we encounter might be applicable in other infrastructural settings, we need to understand better if the training of HCI and health researchers in other countries facilitates better connections and research integration. We only investigated HCI-Health collaborations in university settings. Partnerships between academia and industry are increasingly common. However, such collaborations are subject to different types of incentives and motivations to do cross-disciplinary work as well as different management practices, which can further complicate the collaboration \cite{boules2016future,bozeman2013evidence}. More research is needed to deeply understand these challenges and supports needed in academic-industry collaborations. 

\textit{Understand health collaborator backgrounds and infrastructure.} 
In this research, we do not strictly differentiate between the disciplinary background of health collaborators. Our work is limited by collaborations in which participants developed technologies, typically for behavioral interventions. Health researchers are not a homogeneous group and have many disciplinary backgrounds that might make it easier or difficult to integrate research perspectives. More research is needed to understand how the different training of health researchers integrates with HCI methods. Similarly, the infrastructure of different health researchers might vary. Different health departments might include different job structure, salary models, or security of employment \cite{walling2018academic,mchale2019promotion}. Further research is necessary to understand the role of health related departments on the success of cross-disciplinary collaborations.

\textit{Understand the role of different disciplines and team member roles in HCI collaborations}. The current research investigated challenges that primary investigators encounter when conducting cross-disciplinary research in HCI-Health. However, there is a need to understand the structure of the teams engaged in this type of research as well as the distribution of roles and responsibilities to better support the research collaboration. Further research is needed to understand how research collaborations work or break down when researchers from other disciplines join the team, or when collaborations are formed with external teams such as firms providing programming and design services.

\textit{Investigate further methods.}
As community-based research methods become more prevalent in HCI \cite{harrington2019engaging}, how the HCI field supports the collaboration needs and goals of community-based research team members will be an increasingly important consideration of teamwork. 

This paper emphasizes the experiences at the intersection of two fields: HCI and Health. Yet, researchers at this intersection may also identify as a member of additional related fields, such as Health Informatics and Human Factors. This, in addition to the differing department infrastructures, might affect the type of research contributions people strive to make. Additional research can help us understand how the practices, norms, and standards of related fields can influence the types of challenges researchers face, the strategies they use to navigate such challenges, and what lessons can be adopted across fields to enable more robust collaboration. 

\section{Conclusion}
In this paper, we summarize the results of an interview study with 17 early- and mid-career faculty conducting research at the intersection of HCI and Health. Our findings shed light on the many challenges they face in establishing research agendas, such as finding collaborations and navigating collaborations with misaligned goals. These findings also show how our study participants navigated these challenges, aligned their goals and desired collaboration outcomes with ongoing projects, and positioned the value of their methodological approach to health researchers. We share insights on how this work impacted participants' intellectual contributions and career trajectories. 

In addition to participants' lived experiences, we apply a Team Science lens to reveal what opportunities exist to support these individuals as they collaborate with health researchers, such as leveraging social connections, proximity, joint conceptualization of a problem space, and beneficial career outcomes. We also discuss how best practices from Team Science can be implemented and extended for researchers at the intersection of HCI, including toolkits to align with collaborators on research goals, revisions to tenure, promotion policies to account for the invisible work of establishing collaborations, and changes to funding agency incentives. 

By implementing these recommendations, the HCI community can begin to recognize the challenges early-career faculty must navigate and scaffold the individual and institutional support needed for their success. In taking these steps, we strive towards a transdisciplinary research vision, where we tightly integrate with other research domains, produce innovative solutions, and ensure the work of achieving this vision is sustainable for emerging generations of HCI researchers. 

\section{Acknowledgments}
We thank the participants for their time and generosity in sharing honest and vulnerable experiences. We thank the anonymous reviewers, Madhu Reddy, Gloria Mark, Judy Olson, Anne Marie Piper, Daniel Epstein, Jed Brubaker and Jennifer Turns. Their feedback helped us improve the study and the current paper. 

%% file: 00main.bbl

\begin{thebibliography}{128}

%%% the Web address.

\ifx \showCODEN    \undefined \def \showCODEN     #1{\unskip}     \fi
\ifx \showDOI      \undefined \def \showDOI       #1{#1}\fi
\ifx \showISBNx    \undefined \def \showISBNx     #1{\unskip}     \fi
\ifx \showISBNxiii \undefined \def \showISBNxiii  #1{\unskip}     \fi
\ifx \showISSN     \undefined \def \showISSN      #1{\unskip}     \fi
\ifx \showLCCN     \undefined \def \showLCCN      #1{\unskip}     \fi
\ifx \shownote     \undefined \def \shownote      #1{#1}          \fi
\ifx \showarticletitle \undefined \def \showarticletitle #1{#1}   \fi
\ifx \showURL      \undefined \def \showURL       {\relax}        \fi
% invisible to TeX
\providecommand\bibfield[2]{#2}
\providecommand\bibinfo[2]{#2}
\providecommand\natexlab[1]{#1}
\providecommand\showeprint[2][]{arXiv:#2}

\bibitem[\protect\citeauthoryear{??}{tea}{2022a}]%
        {teamsciencetoolkit-uw-promotion}
 \bibinfo{year}{Retrieved on April 23rd, 2022}\natexlab{a}.
\newblock \bibinfo{title}{Appointment, Promotion and Tenure (APT) Toolkit}.
\newblock
  \bibinfo{howpublished}{\url{"https://collaborate.uw.edu/online-training-and-resources/apt-toolkit/"}}.
\newblock


\bibitem[\protect\citeauthoryear{??}{nci}{2022}]%
        {nci-pathways}
 \bibinfo{year}{Retrieved on April 23rd, 2022}\natexlab{}.
\newblock \bibinfo{title}{National Cancer Institute. (2015). Key Initiatives:
  NCI Network on Biobehavioral Pathways in Cancer}.
\newblock
  \bibinfo{howpublished}{\url{"http://cancercontrol.cancer.gov/brp/bbpsb/ncintwk-biopthwys.html"}}.
\newblock


\bibitem[\protect\citeauthoryear{??}{tea}{2022b}]%
        {teamsciencetoolkit-uw}
 \bibinfo{year}{Retrieved on April 23rd, 2022}\natexlab{b}.
\newblock \bibinfo{title}{Team Science Initiative, University of Washington
  Toolkits}.
\newblock
  \bibinfo{howpublished}{\url{"https://collaborate.uw.edu/programs/team-science-initiative/"}}.
\newblock


\bibitem[\protect\citeauthoryear{??}{nih}{2021a}]%
        {nih-hci}
 \bibinfo{year}{Retrieved on September 8, 2021}\natexlab{a}.
\newblock \bibinfo{title}{NIH grants including HCI approaches}.
\newblock \bibinfo{howpublished}{\url{"
  https://reporter.nih.gov/search/Aac-SUWwXkGnSTGfCqxA4g/projects"}}.
\newblock


\bibitem[\protect\citeauthoryear{??}{nih}{2021b}]%
        {nih-training-center}
 \bibinfo{year}{Retrieved on September 8, 2021}\natexlab{b}.
\newblock \bibinfo{title}{NIH Training and Center Grants Including HCI
  approaches}.
\newblock
  \bibinfo{howpublished}{\url{"https://reporter.nih.gov/search/xesAWWCm_0Wxwn9tbg0BzQ/projects"}}.
\newblock


\bibitem[\protect\citeauthoryear{??}{mht}{2021}]%
        {mhti}
 \bibinfo{year}{Retrieved on September 8th, 2021}\natexlab{}.
\newblock \bibinfo{title}{mHealth Training Institute}.
\newblock \bibinfo{howpublished}{\url{"https://mhti.md2k.org/"}}.
\newblock


\bibitem[\protect\citeauthoryear{??}{nsf}{2021}]%
        {nsf-sch}
 \bibinfo{year}{Retrieved on September 8th, 2021}\natexlab{}.
\newblock \bibinfo{title}{NSF Smart and Connected Health}.
\newblock
  \bibinfo{howpublished}{\url{"https://www.nsf.gov/pubs/2018/nsf18541/nsf18541.htm"}}.
\newblock


\bibitem[\protect\citeauthoryear{??}{tea}{2021}]%
        {teamsciencetoolkit}
 \bibinfo{year}{Retrieved on September 8th, 2021}\natexlab{}.
\newblock \bibinfo{title}{Team Science Toolkit}.
\newblock
  \bibinfo{howpublished}{\url{"https://www.teamscience.net/modules/module1-resources"}}.
\newblock


\bibitem[\protect\citeauthoryear{Abbott, MacLeod, Nurain, Ekobe, and
  Patil}{Abbott et~al\mbox{.}}{2019}]%
        {abbott2019local}
\bibfield{author}{\bibinfo{person}{Jacob Abbott}, \bibinfo{person}{Haley
  MacLeod}, \bibinfo{person}{Novia Nurain}, \bibinfo{person}{Gustave Ekobe},
  {and} \bibinfo{person}{Sameer Patil}.} \bibinfo{year}{2019}\natexlab{}.
\newblock \showarticletitle{Local standards for anonymization practices in
  health, wellness, accessibility, and aging research at CHI}. In
  \bibinfo{booktitle}{\emph{Proceedings of the 2019 CHI Conference on Human
  Factors in Computing Systems}}. \bibinfo{pages}{1--14}.
\newblock


\bibitem[\protect\citeauthoryear{Adler and Chen}{Adler and Chen}{2011}]%
        {adler2011combining}
\bibfield{author}{\bibinfo{person}{Paul~S Adler} {and}
  \bibinfo{person}{Clara~Xiaoling Chen}.} \bibinfo{year}{2011}\natexlab{}.
\newblock \showarticletitle{Combining creativity and control: Understanding
  individual motivation in large-scale collaborative creativity}.
\newblock \bibinfo{journal}{\emph{Accounting, organizations and society}}
  \bibinfo{volume}{36}, \bibinfo{number}{2} (\bibinfo{year}{2011}),
  \bibinfo{pages}{63--85}.
\newblock


\bibitem[\protect\citeauthoryear{Aggarwal, Hoang, Ploderer, Vetere, Khot, and
  Bradford}{Aggarwal et~al\mbox{.}}{2020}]%
        {aggarwal2020lessons}
\bibfield{author}{\bibinfo{person}{Deepti Aggarwal}, \bibinfo{person}{Thuong
  Hoang}, \bibinfo{person}{Bernd Ploderer}, \bibinfo{person}{Frank Vetere},
  \bibinfo{person}{Rohit~Ashok Khot}, {and} \bibinfo{person}{Mark Bradford}.}
  \bibinfo{year}{2020}\natexlab{}.
\newblock \showarticletitle{Lessons Learnt from Designing a Smart Clothing
  Telehealth System for Hospital Use}. In \bibinfo{booktitle}{\emph{32nd
  Australian Conference on Human-Computer Interaction}}.
  \bibinfo{pages}{355--367}.
\newblock


\bibitem[\protect\citeauthoryear{Ancker, Benda, Reddy, Unertl, and
  Veinot}{Ancker et~al\mbox{.}}{2021}]%
        {ancker2021guidance}
\bibfield{author}{\bibinfo{person}{Jessica~S Ancker},
  \bibinfo{person}{Natalie~C Benda}, \bibinfo{person}{Madhu Reddy},
  \bibinfo{person}{Kim~M Unertl}, {and} \bibinfo{person}{Tiffany Veinot}.}
  \bibinfo{year}{2021}\natexlab{}.
\newblock \showarticletitle{Guidance for publishing qualitative research in
  informatics}.
\newblock \bibinfo{journal}{\emph{Journal of the American Medical Informatics
  Association}} \bibinfo{volume}{28}, \bibinfo{number}{12}
  (\bibinfo{year}{2021}), \bibinfo{pages}{2743--2748}.
\newblock


\bibitem[\protect\citeauthoryear{Balakrishnan, Kiesler, Cummings, and
  Zadeh}{Balakrishnan et~al\mbox{.}}{2011}]%
        {balakrishnan2011research}
\bibfield{author}{\bibinfo{person}{Aruna~D Balakrishnan}, \bibinfo{person}{Sara
  Kiesler}, \bibinfo{person}{Jonathon~N Cummings}, {and} \bibinfo{person}{Reza
  Zadeh}.} \bibinfo{year}{2011}\natexlab{}.
\newblock \showarticletitle{Research team integration: What it is and why it
  matters}. In \bibinfo{booktitle}{\emph{Proceedings of the ACM 2011 conference
  on Computer supported cooperative work}}. \bibinfo{pages}{523--532}.
\newblock


\bibitem[\protect\citeauthoryear{Beede, Baylor, Hersch, Iurchenko, Wilcox,
  Ruamviboonsuk, and Vardoulakis}{Beede et~al\mbox{.}}{2020}]%
        {beede2020human}
\bibfield{author}{\bibinfo{person}{Emma Beede}, \bibinfo{person}{Elizabeth
  Baylor}, \bibinfo{person}{Fred Hersch}, \bibinfo{person}{Anna Iurchenko},
  \bibinfo{person}{Lauren Wilcox}, \bibinfo{person}{Paisan Ruamviboonsuk},
  {and} \bibinfo{person}{Laura~M Vardoulakis}.}
  \bibinfo{year}{2020}\natexlab{}.
\newblock \showarticletitle{A human-centered evaluation of a deep learning
  system deployed in clinics for the detection of diabetic retinopathy}. In
  \bibinfo{booktitle}{\emph{Proceedings of the 2020 CHI Conference on Human
  Factors in Computing Systems}}. \bibinfo{pages}{1--12}.
\newblock


\bibitem[\protect\citeauthoryear{Bietz, Abrams, Cooper, Stevens, Puga, Patel,
  Olson, and Olson}{Bietz et~al\mbox{.}}{2012a}]%
        {bietz2012improving}
\bibfield{author}{\bibinfo{person}{Matthew~J Bietz}, \bibinfo{person}{Steve
  Abrams}, \bibinfo{person}{Dan~M Cooper}, \bibinfo{person}{Kathleen~R
  Stevens}, \bibinfo{person}{Frank Puga}, \bibinfo{person}{Darpan~I Patel},
  \bibinfo{person}{Gary~M Olson}, {and} \bibinfo{person}{Judith~S Olson}.}
  \bibinfo{year}{2012}\natexlab{a}.
\newblock \showarticletitle{Improving the odds through the Collaboration
  Success Wizard}.
\newblock \bibinfo{journal}{\emph{Translational behavioral medicine}}
  \bibinfo{volume}{2}, \bibinfo{number}{4} (\bibinfo{year}{2012}),
  \bibinfo{pages}{480--486}.
\newblock


\bibitem[\protect\citeauthoryear{Bietz, Ferro, and Lee}{Bietz
  et~al\mbox{.}}{2012b}]%
        {bietz2012sustaining}
\bibfield{author}{\bibinfo{person}{Matthew~J Bietz}, \bibinfo{person}{Toni
  Ferro}, {and} \bibinfo{person}{Charlotte~P Lee}.}
  \bibinfo{year}{2012}\natexlab{b}.
\newblock \showarticletitle{Sustaining the development of cyberinfrastructure:
  an organization adapting to change}. In \bibinfo{booktitle}{\emph{Proceedings
  of the ACM 2012 conference on Computer Supported Cooperative Work}}.
  \bibinfo{pages}{901--910}.
\newblock


\bibitem[\protect\citeauthoryear{Bietz and Lee}{Bietz and Lee}{2012}]%
        {bietz2012adapting}
\bibfield{author}{\bibinfo{person}{Matthew~J Bietz} {and}
  \bibinfo{person}{Charlotte~P Lee}.} \bibinfo{year}{2012}\natexlab{}.
\newblock \showarticletitle{Adapting cyberinfrastructure to new science:
  tensions and strategies}.
\newblock In \bibinfo{booktitle}{\emph{Proceedings of the 2012 iConference}}.
  \bibinfo{pages}{183--190}.
\newblock


\bibitem[\protect\citeauthoryear{Binz-Scharf, Kalish, and Paik}{Binz-Scharf
  et~al\mbox{.}}{2015}]%
        {binz2015making}
\bibfield{author}{\bibinfo{person}{Maria~C Binz-Scharf}, \bibinfo{person}{Yuval
  Kalish}, {and} \bibinfo{person}{Leslie Paik}.}
  \bibinfo{year}{2015}\natexlab{}.
\newblock \showarticletitle{Making science: New generations of collaborative
  knowledge production}.
\newblock \bibinfo{journal}{\emph{American Behavioral Scientist}}
  \bibinfo{volume}{59}, \bibinfo{number}{5} (\bibinfo{year}{2015}),
  \bibinfo{pages}{531--547}.
\newblock


\bibitem[\protect\citeauthoryear{Blandford}{Blandford}{2018}]%
        {blandford2018lessons}
\bibfield{author}{\bibinfo{person}{Ann Blandford}.}
  \bibinfo{year}{2018}\natexlab{}.
\newblock \showarticletitle{Lessons from working with researchers and
  practitioners in healthcare}.
\newblock \bibinfo{journal}{\emph{Interactions}} \bibinfo{volume}{26},
  \bibinfo{number}{1} (\bibinfo{year}{2018}), \bibinfo{pages}{72--75}.
\newblock


\bibitem[\protect\citeauthoryear{Blandford}{Blandford}{2019}]%
        {blandford2019hci4health}
\bibfield{author}{\bibinfo{person}{Ann Blandford}.}
  \bibinfo{year}{2019}\natexlab{}.
\newblock \showarticletitle{HCI for health and wellbeing: challenges and
  opportunities}.
\newblock \bibinfo{journal}{\emph{International journal of human-computer
  studies}}  \bibinfo{volume}{131} (\bibinfo{year}{2019}),
  \bibinfo{pages}{41--51}.
\newblock


\bibitem[\protect\citeauthoryear{Blandford, Berndt, Catchpole, Furniss, Mayer,
  Mentis, O’kane, Owen, Rajkomar, and Randell}{Blandford
  et~al\mbox{.}}{2015}]%
        {blandford2015strategies}
\bibfield{author}{\bibinfo{person}{Ann Blandford}, \bibinfo{person}{Erik
  Berndt}, \bibinfo{person}{Ken Catchpole}, \bibinfo{person}{Dominic Furniss},
  \bibinfo{person}{Astrid Mayer}, \bibinfo{person}{Helena Mentis},
  \bibinfo{person}{Aisling~Ann O’kane}, \bibinfo{person}{Tom Owen},
  \bibinfo{person}{Atish Rajkomar}, {and} \bibinfo{person}{Rebecca Randell}.}
  \bibinfo{year}{2015}\natexlab{}.
\newblock \showarticletitle{Strategies for conducting situated studies of
  technology use in hospitals}.
\newblock \bibinfo{journal}{\emph{Cognition, technology \& work}}
  \bibinfo{volume}{17}, \bibinfo{number}{4} (\bibinfo{year}{2015}),
  \bibinfo{pages}{489--502}.
\newblock


\bibitem[\protect\citeauthoryear{Blandford, Gibbs, Newhouse, Perski, Singh, and
  Murray}{Blandford et~al\mbox{.}}{2018}]%
        {blandford2018seven}
\bibfield{author}{\bibinfo{person}{Ann Blandford}, \bibinfo{person}{Jo Gibbs},
  \bibinfo{person}{Nikki Newhouse}, \bibinfo{person}{Olga Perski},
  \bibinfo{person}{Aneesha Singh}, {and} \bibinfo{person}{Elizabeth Murray}.}
  \bibinfo{year}{2018}\natexlab{}.
\newblock \showarticletitle{Seven lessons for interdisciplinary research on
  interactive digital health interventions}.
\newblock \bibinfo{journal}{\emph{Digital health}}  \bibinfo{volume}{4}
  (\bibinfo{year}{2018}), \bibinfo{pages}{2055207618770325}.
\newblock


\bibitem[\protect\citeauthoryear{Boules, Douglas, Feldman, Fix, Hager,
  Hailpern, Hebert, Lopresti, Mynatt, Rossbach, et~al\mbox{.}}{Boules
  et~al\mbox{.}}{2016}]%
        {boules2016future}
\bibfield{author}{\bibinfo{person}{Nady Boules}, \bibinfo{person}{Khari
  Douglas}, \bibinfo{person}{Stuart Feldman}, \bibinfo{person}{Limor Fix},
  \bibinfo{person}{Gregory Hager}, \bibinfo{person}{Brent Hailpern},
  \bibinfo{person}{Martial Hebert}, \bibinfo{person}{Dan Lopresti},
  \bibinfo{person}{Beth Mynatt}, \bibinfo{person}{Chris Rossbach},
  {et~al\mbox{.}}} \bibinfo{year}{2016}\natexlab{}.
\newblock \showarticletitle{The future of computing research: industry-academic
  collaborations}.
\newblock \bibinfo{journal}{\emph{arXiv preprint arXiv:1606.09236}}
  (\bibinfo{year}{2016}).
\newblock


\bibitem[\protect\citeauthoryear{Bozeman and Boardman}{Bozeman and
  Boardman}{2013}]%
        {bozeman2013evidence}
\bibfield{author}{\bibinfo{person}{Barry Bozeman} {and} \bibinfo{person}{Craig
  Boardman}.} \bibinfo{year}{2013}\natexlab{}.
\newblock \showarticletitle{An evidence-based assessment of research
  collaboration and team science: patterns in industry and university-industry
  partnerships}. In \bibinfo{booktitle}{\emph{Workshop on Institutional and
  Organizational Supports for Team Science}}.
\newblock


\bibitem[\protect\citeauthoryear{Buis and Huh-Yoo}{Buis and Huh-Yoo}{2020}]%
        {buis2020common}
\bibfield{author}{\bibinfo{person}{Lorraine~R Buis} {and} \bibinfo{person}{Jina
  Huh-Yoo}.} \bibinfo{year}{2020}\natexlab{}.
\newblock \showarticletitle{Common Shortcomings in Applying User-Centered
  Design for Digital Health}.
\newblock \bibinfo{journal}{\emph{IEEE Pervasive Computing}}
  \bibinfo{volume}{19}, \bibinfo{number}{3} (\bibinfo{year}{2020}),
  \bibinfo{pages}{45--49}.
\newblock


\bibitem[\protect\citeauthoryear{Cai, Winter, Steiner, Wilcox, and Terry}{Cai
  et~al\mbox{.}}{2019}]%
        {cai2019hello}
\bibfield{author}{\bibinfo{person}{Carrie~J Cai}, \bibinfo{person}{Samantha
  Winter}, \bibinfo{person}{David Steiner}, \bibinfo{person}{Lauren Wilcox},
  {and} \bibinfo{person}{Michael Terry}.} \bibinfo{year}{2019}\natexlab{}.
\newblock \showarticletitle{" Hello AI": Uncovering the Onboarding Needs of
  Medical Practitioners for Human-AI Collaborative Decision-Making}.
\newblock \bibinfo{journal}{\emph{Proceedings of the ACM on Human-computer
  Interaction}} \bibinfo{volume}{3}, \bibinfo{number}{CSCW}
  (\bibinfo{year}{2019}), \bibinfo{pages}{1--24}.
\newblock


\bibitem[\protect\citeauthoryear{Cajander and Gr{\"u}nloh}{Cajander and
  Gr{\"u}nloh}{2019}]%
        {cajander2019electronic}
\bibfield{author}{\bibinfo{person}{{\AA}sa Cajander} {and}
  \bibinfo{person}{Christiane Gr{\"u}nloh}.} \bibinfo{year}{2019}\natexlab{}.
\newblock \showarticletitle{Electronic health records are more than a work
  tool: conflicting needs of direct and indirect stakeholders}. In
  \bibinfo{booktitle}{\emph{Proceedings of the 2019 CHI Conference on Human
  Factors in Computing Systems}}. \bibinfo{pages}{1--13}.
\newblock


\bibitem[\protect\citeauthoryear{Campbell}{Campbell}{2005}]%
        {campbell2005overcoming}
\bibfield{author}{\bibinfo{person}{Lisa~M Campbell}.}
  \bibinfo{year}{2005}\natexlab{}.
\newblock \showarticletitle{Overcoming obstacles to interdisciplinary
  research}.
\newblock \bibinfo{journal}{\emph{Conservation biology}} \bibinfo{volume}{19},
  \bibinfo{number}{2} (\bibinfo{year}{2005}), \bibinfo{pages}{574--577}.
\newblock


\bibitem[\protect\citeauthoryear{Chung, Dew, Cole, Zia, Fogarty, Kientz, and
  Munson}{Chung et~al\mbox{.}}{2016}]%
        {chung2016boundary}
\bibfield{author}{\bibinfo{person}{Chia-Fang Chung}, \bibinfo{person}{Kristin
  Dew}, \bibinfo{person}{Allison Cole}, \bibinfo{person}{Jasmine Zia},
  \bibinfo{person}{James Fogarty}, \bibinfo{person}{Julie~A Kientz}, {and}
  \bibinfo{person}{Sean~A Munson}.} \bibinfo{year}{2016}\natexlab{}.
\newblock \showarticletitle{Boundary negotiating artifacts in personal
  informatics: patient-provider collaboration with patient-generated data}. In
  \bibinfo{booktitle}{\emph{Proceedings of the 19th ACM Conference on
  Computer-Supported Cooperative Work \& Social Computing}}.
  \bibinfo{pages}{770--786}.
\newblock


\bibitem[\protect\citeauthoryear{Cibrian, Monteiro, Schuck, Nelson, Hayes, and
  Lakes}{Cibrian et~al\mbox{.}}{2022}]%
        {cibrian2022interdisciplinary}
\bibfield{author}{\bibinfo{person}{Franceli~L Cibrian}, \bibinfo{person}{Elissa
  Monteiro}, \bibinfo{person}{Sabrina~EB Schuck}, \bibinfo{person}{Michele
  Nelson}, \bibinfo{person}{Gillian~R Hayes}, {and}
  \bibinfo{person}{Kimberley~D Lakes}.} \bibinfo{year}{2022}\natexlab{}.
\newblock \showarticletitle{Interdisciplinary Tensions When Developing Digital
  Interventions Supporting Individuals With ADHD}.
\newblock \bibinfo{journal}{\emph{Frontiers in Digital Health}}
  \bibinfo{volume}{4} (\bibinfo{year}{2022}).
\newblock


\bibitem[\protect\citeauthoryear{Colusso, Bennett, Hsieh, and Munson}{Colusso
  et~al\mbox{.}}{2017}]%
        {colusso2017translational}
\bibfield{author}{\bibinfo{person}{Lucas Colusso}, \bibinfo{person}{Cynthia~L
  Bennett}, \bibinfo{person}{Gary Hsieh}, {and} \bibinfo{person}{Sean~A
  Munson}.} \bibinfo{year}{2017}\natexlab{}.
\newblock \showarticletitle{Translational resources: Reducing the gap between
  academic research and HCI practice}. In \bibinfo{booktitle}{\emph{Proceedings
  of the 2017 Conference on Designing Interactive Systems}}.
  \bibinfo{pages}{957--968}.
\newblock


\bibitem[\protect\citeauthoryear{Colusso, Do, and Hsieh}{Colusso
  et~al\mbox{.}}{2018}]%
        {colusso2018behavior}
\bibfield{author}{\bibinfo{person}{Lucas Colusso}, \bibinfo{person}{Tien Do},
  {and} \bibinfo{person}{Gary Hsieh}.} \bibinfo{year}{2018}\natexlab{}.
\newblock \showarticletitle{Behavior change design sprints}. In
  \bibinfo{booktitle}{\emph{Proceedings of the 2018 Designing Interactive
  Systems Conference}}. \bibinfo{pages}{791--803}.
\newblock


\bibitem[\protect\citeauthoryear{Colusso, Jones, Munson, and Hsieh}{Colusso
  et~al\mbox{.}}{2019}]%
        {colusso2019translational}
\bibfield{author}{\bibinfo{person}{Lucas Colusso}, \bibinfo{person}{Ridley
  Jones}, \bibinfo{person}{Sean~A Munson}, {and} \bibinfo{person}{Gary Hsieh}.}
  \bibinfo{year}{2019}\natexlab{}.
\newblock \showarticletitle{A translational science model for HCI}. In
  \bibinfo{booktitle}{\emph{Proceedings of the 2019 CHI Conference on Human
  Factors in Computing Systems}}. \bibinfo{pages}{1--13}.
\newblock


\bibitem[\protect\citeauthoryear{Council et~al\mbox{.}}{Council
  et~al\mbox{.}}{2014}]%
        {national2014convergence}
\bibfield{author}{\bibinfo{person}{National~Research Council} {et~al\mbox{.}}}
  \bibinfo{year}{2014}\natexlab{}.
\newblock \bibinfo{booktitle}{\emph{Convergence: Facilitating transdisciplinary
  integration of life sciences, physical sciences, engineering, and beyond}}.
\newblock \bibinfo{publisher}{National Academies Press}.
\newblock


\bibitem[\protect\citeauthoryear{Council et~al\mbox{.}}{Council
  et~al\mbox{.}}{2015}]%
        {national2015enhancing}
\bibfield{author}{\bibinfo{person}{National~Research Council} {et~al\mbox{.}}}
  \bibinfo{year}{2015}\natexlab{}.
\newblock \showarticletitle{Enhancing the effectiveness of team science}.
\newblock  (\bibinfo{year}{2015}).
\newblock


\bibitem[\protect\citeauthoryear{Cukier, Klyachko, and Spring}{Cukier
  et~al\mbox{.}}{2020}]%
        {cukier2020team}
\bibfield{author}{\bibinfo{person}{Sasha Cukier}, \bibinfo{person}{Ekaterina
  Klyachko}, {and} \bibinfo{person}{Bonnie Spring}.}
  \bibinfo{year}{2020}\natexlab{}.
\newblock \showarticletitle{Team Science in Health}.
\newblock \bibinfo{journal}{\emph{The Wiley Encyclopedia of Health Psychology}}
  (\bibinfo{year}{2020}), \bibinfo{pages}{243--252}.
\newblock


\bibitem[\protect\citeauthoryear{Doherty, Marcano-Belisario, Cohn, Mastellos,
  Morrison, Car, and Doherty}{Doherty et~al\mbox{.}}{2019}]%
        {doherty2019engagement}
\bibfield{author}{\bibinfo{person}{Kevin Doherty}, \bibinfo{person}{Jos{\'e}
  Marcano-Belisario}, \bibinfo{person}{Martin Cohn}, \bibinfo{person}{Nikolaos
  Mastellos}, \bibinfo{person}{Cecily Morrison}, \bibinfo{person}{Josip Car},
  {and} \bibinfo{person}{Gavin Doherty}.} \bibinfo{year}{2019}\natexlab{}.
\newblock \showarticletitle{Engagement with mental health screening on mobile
  devices: Results from an antenatal feasibility study}. In
  \bibinfo{booktitle}{\emph{Proceedings of the 2019 CHI Conference on Human
  Factors in Computing Systems}}. \bibinfo{pages}{1--15}.
\newblock


\bibitem[\protect\citeauthoryear{Dopp, Parisi, Munson, and Lyon}{Dopp
  et~al\mbox{.}}{2019}]%
        {dopp2019glossary}
\bibfield{author}{\bibinfo{person}{Alex~R Dopp}, \bibinfo{person}{Kathryn~E
  Parisi}, \bibinfo{person}{Sean~A Munson}, {and} \bibinfo{person}{Aaron~R
  Lyon}.} \bibinfo{year}{2019}\natexlab{}.
\newblock \showarticletitle{A glossary of user-centered design strategies for
  implementation experts}.
\newblock \bibinfo{journal}{\emph{Translational behavioral medicine}}
  \bibinfo{volume}{9}, \bibinfo{number}{6} (\bibinfo{year}{2019}),
  \bibinfo{pages}{1057--1064}.
\newblock


\bibitem[\protect\citeauthoryear{Edwards and Moczygemba}{Edwards and
  Moczygemba}{2004}]%
        {edwards2004reducing}
\bibfield{author}{\bibinfo{person}{Marie Edwards} {and} \bibinfo{person}{Jackie
  Moczygemba}.} \bibinfo{year}{2004}\natexlab{}.
\newblock \showarticletitle{Reducing medical errors through better
  documentation}.
\newblock \bibinfo{journal}{\emph{The health care manager}}
  \bibinfo{volume}{23}, \bibinfo{number}{4} (\bibinfo{year}{2004}),
  \bibinfo{pages}{329--333}.
\newblock


\bibitem[\protect\citeauthoryear{Fiore and Schooler}{Fiore and
  Schooler}{2004}]%
        {fiore2004process}
\bibfield{author}{\bibinfo{person}{Stephen~M Fiore} {and}
  \bibinfo{person}{Jonathan~W Schooler}.} \bibinfo{year}{2004}\natexlab{}.
\newblock \showarticletitle{Process mapping and shared cognition: Teamwork and
  the development of shared problem models.}
\newblock  (\bibinfo{year}{2004}).
\newblock


\bibitem[\protect\citeauthoryear{Frodeman, Klein, and Pacheco}{Frodeman
  et~al\mbox{.}}{2017}]%
        {frodeman2017oxford}
\bibfield{author}{\bibinfo{person}{Robert Frodeman},
  \bibinfo{person}{Julie~Thompson Klein}, {and} \bibinfo{person}{Roberto Carlos
  Dos~Santos Pacheco}.} \bibinfo{year}{2017}\natexlab{}.
\newblock \bibinfo{booktitle}{\emph{The Oxford handbook of
  interdisciplinarity}}.
\newblock \bibinfo{publisher}{Oxford University Press}.
\newblock


\bibitem[\protect\citeauthoryear{Furniss, Randell, O’kane, Taneva, Mentis,
  and Blandford}{Furniss et~al\mbox{.}}{2014}]%
        {furniss2014fieldwork}
\bibfield{author}{\bibinfo{person}{Dominic Furniss}, \bibinfo{person}{Rebecca
  Randell}, \bibinfo{person}{Aisling~Ann O’kane}, \bibinfo{person}{Svetlena
  Taneva}, \bibinfo{person}{Helena Mentis}, {and} \bibinfo{person}{Ann
  Blandford}.} \bibinfo{year}{2014}\natexlab{}.
\newblock \showarticletitle{Fieldwork for healthcare: guidance for
  investigating human factors in computing systems}.
\newblock \bibinfo{journal}{\emph{Synthesis Lectures on Assistive,
  Rehabilitative, and Health-Preserving Technologies}} \bibinfo{volume}{3},
  \bibinfo{number}{2} (\bibinfo{year}{2014}), \bibinfo{pages}{1--146}.
\newblock


\bibitem[\protect\citeauthoryear{Glied, Bakken, Formicola, Gebbie, and
  Larson}{Glied et~al\mbox{.}}{2007}]%
        {glied2007institutional}
\bibfield{author}{\bibinfo{person}{Sherry Glied}, \bibinfo{person}{Suzanne
  Bakken}, \bibinfo{person}{Allan Formicola}, \bibinfo{person}{Kristine
  Gebbie}, {and} \bibinfo{person}{Elaine~L Larson}.}
  \bibinfo{year}{2007}\natexlab{}.
\newblock \showarticletitle{Institutional challenges of interdisciplinary
  research centers.}
\newblock \bibinfo{journal}{\emph{Journal of Research Administration}}
  \bibinfo{volume}{38}, \bibinfo{number}{2} (\bibinfo{year}{2007}),
  \bibinfo{pages}{28--36}.
\newblock


\bibitem[\protect\citeauthoryear{Hackman}{Hackman}{2012}]%
        {hackman2012causes}
\bibfield{author}{\bibinfo{person}{J~Richard Hackman}.}
  \bibinfo{year}{2012}\natexlab{}.
\newblock \showarticletitle{From causes to conditions in group research}.
\newblock \bibinfo{journal}{\emph{Journal of organizational Behavior}}
  \bibinfo{volume}{33}, \bibinfo{number}{3} (\bibinfo{year}{2012}),
  \bibinfo{pages}{428--444}.
\newblock


\bibitem[\protect\citeauthoryear{Haldar, Kim, Mishra, Hartzler, Pollack, and
  Pratt}{Haldar et~al\mbox{.}}{2020}]%
        {haldar2020patient}
\bibfield{author}{\bibinfo{person}{Shefali Haldar}, \bibinfo{person}{Yoojung
  Kim}, \bibinfo{person}{Sonali~R Mishra}, \bibinfo{person}{Andrea~L Hartzler},
  \bibinfo{person}{Ari~H Pollack}, {and} \bibinfo{person}{Wanda Pratt}.}
  \bibinfo{year}{2020}\natexlab{}.
\newblock \showarticletitle{The Patient Advice System: A Technology Probe Study
  to Enable Peer Support in the Hospital}.
\newblock \bibinfo{journal}{\emph{Proceedings of the ACM on Human-Computer
  Interaction}} \bibinfo{volume}{4}, \bibinfo{number}{CSCW2}
  (\bibinfo{year}{2020}), \bibinfo{pages}{1--23}.
\newblock


\bibitem[\protect\citeauthoryear{Hall, Vogel, Ku, Klein, Banacki, Bennett,
  et~al\mbox{.}}{Hall et~al\mbox{.}}{2013}]%
        {hall2013recognition}
\bibfield{author}{\bibinfo{person}{KL Hall}, \bibinfo{person}{AL Vogel},
  \bibinfo{person}{MC Ku}, \bibinfo{person}{JT Klein}, \bibinfo{person}{A
  Banacki}, \bibinfo{person}{LM Bennett}, {et~al\mbox{.}}}
  \bibinfo{year}{2013}\natexlab{}.
\newblock \showarticletitle{Recognition for team science and
  cross-disciplinarity in academia: an exploration of Promotion and Tenure
  Policy and Guideline Language from Clinical and Translational Science Awards
  (CTSA) Institutions}. In \bibinfo{booktitle}{\emph{National Academies
  Workshop on Institutional and Organizational Support for Team Science}}.
\newblock


\bibitem[\protect\citeauthoryear{Hall, Vogel, and Crowston}{Hall
  et~al\mbox{.}}{2019}]%
        {hall2019comprehensive}
\bibfield{author}{\bibinfo{person}{Kara~L Hall}, \bibinfo{person}{Amanda~L
  Vogel}, {and} \bibinfo{person}{Kevin Crowston}.}
  \bibinfo{year}{2019}\natexlab{}.
\newblock \showarticletitle{Comprehensive collaboration plans: practical
  considerations spanning across individual collaborators to institutional
  supports}.
\newblock In \bibinfo{booktitle}{\emph{Strategies for Team Science Success}}.
  \bibinfo{publisher}{Springer}, \bibinfo{pages}{587--612}.
\newblock


\bibitem[\protect\citeauthoryear{Hall, Vogel, Huang, Serrano, Rice,
  Tsakraklides, and Fiore}{Hall et~al\mbox{.}}{2018}]%
        {hall2018science}
\bibfield{author}{\bibinfo{person}{Kara~L Hall}, \bibinfo{person}{Amanda~L
  Vogel}, \bibinfo{person}{Grace~C Huang}, \bibinfo{person}{Katrina~J Serrano},
  \bibinfo{person}{Elise~L Rice}, \bibinfo{person}{Sophia~P Tsakraklides},
  {and} \bibinfo{person}{Stephen~M Fiore}.} \bibinfo{year}{2018}\natexlab{}.
\newblock \showarticletitle{The science of team science: A review of the
  empirical evidence and research gaps on collaboration in science.}
\newblock \bibinfo{journal}{\emph{American Psychologist}} \bibinfo{volume}{73},
  \bibinfo{number}{4} (\bibinfo{year}{2018}), \bibinfo{pages}{532}.
\newblock


\bibitem[\protect\citeauthoryear{Hall, Vogel, Stipelman, Stokols, Morgan, and
  Gehlert}{Hall et~al\mbox{.}}{2012}]%
        {hall2012four}
\bibfield{author}{\bibinfo{person}{Kara~L Hall}, \bibinfo{person}{Amanda~L
  Vogel}, \bibinfo{person}{Brooke~A Stipelman}, \bibinfo{person}{Daniel
  Stokols}, \bibinfo{person}{Glen Morgan}, {and} \bibinfo{person}{Sarah
  Gehlert}.} \bibinfo{year}{2012}\natexlab{}.
\newblock \showarticletitle{A four-phase model of transdisciplinary team-based
  research: goals, team processes, and strategies}.
\newblock \bibinfo{journal}{\emph{Translational behavioral medicine}}
  \bibinfo{volume}{2}, \bibinfo{number}{4} (\bibinfo{year}{2012}),
  \bibinfo{pages}{415--430}.
\newblock


\bibitem[\protect\citeauthoryear{Harper}{Harper}{2005}]%
        {harper2005review}
\bibfield{author}{\bibinfo{person}{Paul~R Harper}.}
  \bibinfo{year}{2005}\natexlab{}.
\newblock \showarticletitle{A review and comparison of classification
  algorithms for medical decision making}.
\newblock \bibinfo{journal}{\emph{Health policy}} \bibinfo{volume}{71},
  \bibinfo{number}{3} (\bibinfo{year}{2005}), \bibinfo{pages}{315--331}.
\newblock


\bibitem[\protect\citeauthoryear{Harrington, Borgos-Rodriguez, and
  Piper}{Harrington et~al\mbox{.}}{2019}]%
        {harrington2019engaging}
\bibfield{author}{\bibinfo{person}{Christina~N Harrington},
  \bibinfo{person}{Katya Borgos-Rodriguez}, {and} \bibinfo{person}{Anne~Marie
  Piper}.} \bibinfo{year}{2019}\natexlab{}.
\newblock \showarticletitle{Engaging low-income African American older adults
  in health discussions through community-based design workshops}. In
  \bibinfo{booktitle}{\emph{Proceedings of the 2019 chi conference on human
  factors in computing systems}}. \bibinfo{pages}{1--15}.
\newblock


\bibitem[\protect\citeauthoryear{Hayes}{Hayes}{2011}]%
        {hayes2011relationship}
\bibfield{author}{\bibinfo{person}{Gillian~R Hayes}.}
  \bibinfo{year}{2011}\natexlab{}.
\newblock \showarticletitle{The relationship of action research to
  human-computer interaction}.
\newblock \bibinfo{journal}{\emph{ACM Transactions on Computer-Human
  Interaction (TOCHI)}} \bibinfo{volume}{18}, \bibinfo{number}{3}
  (\bibinfo{year}{2011}), \bibinfo{pages}{1--20}.
\newblock


\bibitem[\protect\citeauthoryear{Hofmann, Kasnitz, Mankoff, and
  Bennett}{Hofmann et~al\mbox{.}}{2020}]%
        {hofmann2020living}
\bibfield{author}{\bibinfo{person}{Megan Hofmann}, \bibinfo{person}{Devva
  Kasnitz}, \bibinfo{person}{Jennifer Mankoff}, {and}
  \bibinfo{person}{Cynthia~L Bennett}.} \bibinfo{year}{2020}\natexlab{}.
\newblock \showarticletitle{Living disability theory: Reflections on access,
  research, and design}. In \bibinfo{booktitle}{\emph{The 22nd International
  ACM SIGACCESS Conference on Computers and Accessibility}}.
  \bibinfo{pages}{1--13}.
\newblock


\bibitem[\protect\citeauthoryear{Hong, Feustel, Agnihotri, Silverman,
  Simoneaux, and Wilcox}{Hong et~al\mbox{.}}{2017}]%
        {hong2017supporting}
\bibfield{author}{\bibinfo{person}{Matthew~K Hong}, \bibinfo{person}{Clayton
  Feustel}, \bibinfo{person}{Meeshu Agnihotri}, \bibinfo{person}{Max
  Silverman}, \bibinfo{person}{Stephen~F Simoneaux}, {and}
  \bibinfo{person}{Lauren Wilcox}.} \bibinfo{year}{2017}\natexlab{}.
\newblock \showarticletitle{Supporting families in reviewing and communicating
  about radiology imaging studies}. In \bibinfo{booktitle}{\emph{Proceedings of
  the 2017 CHI Conference on Human Factors in Computing Systems}}.
  \bibinfo{pages}{5245--5256}.
\newblock


\bibitem[\protect\citeauthoryear{Huh and Pratt}{Huh and Pratt}{2014}]%
        {huh2014weaving}
\bibfield{author}{\bibinfo{person}{Jina Huh} {and} \bibinfo{person}{Wanda
  Pratt}.} \bibinfo{year}{2014}\natexlab{}.
\newblock \showarticletitle{Weaving clinical expertise in online health
  communities}. In \bibinfo{booktitle}{\emph{Proceedings of the SIGCHI
  Conference on Human Factors in Computing Systems}}.
  \bibinfo{pages}{1355--1364}.
\newblock


\bibitem[\protect\citeauthoryear{Huh-Yoo and Rader}{Huh-Yoo and Rader}{2020}]%
        {huh2020s}
\bibfield{author}{\bibinfo{person}{Jina Huh-Yoo} {and} \bibinfo{person}{Emilee
  Rader}.} \bibinfo{year}{2020}\natexlab{}.
\newblock \showarticletitle{It's the Wild, Wild West: Lessons Learned From IRB
  Members' Risk Perceptions Toward Digital Research Data}.
\newblock \bibinfo{journal}{\emph{Proceedings of the ACM on Human-Computer
  Interaction}} \bibinfo{volume}{4}, \bibinfo{number}{CSCW1}
  (\bibinfo{year}{2020}), \bibinfo{pages}{1--22}.
\newblock


\bibitem[\protect\citeauthoryear{Jacobs, He, F.~Pradier, Lam, Ahn, McCoy,
  Perlis, Doshi-Velez, and Gajos}{Jacobs et~al\mbox{.}}{2021}]%
        {jacobs2021designing}
\bibfield{author}{\bibinfo{person}{Maia Jacobs}, \bibinfo{person}{Jeffrey He},
  \bibinfo{person}{Melanie F.~Pradier}, \bibinfo{person}{Barbara Lam},
  \bibinfo{person}{Andrew~C Ahn}, \bibinfo{person}{Thomas~H McCoy},
  \bibinfo{person}{Roy~H Perlis}, \bibinfo{person}{Finale Doshi-Velez}, {and}
  \bibinfo{person}{Krzysztof~Z Gajos}.} \bibinfo{year}{2021}\natexlab{}.
\newblock \showarticletitle{Designing AI for Trust and Collaboration in
  Time-Constrained Medical Decisions: A Sociotechnical Lens}. In
  \bibinfo{booktitle}{\emph{Proceedings of the 2021 CHI Conference on Human
  Factors in Computing Systems}}. \bibinfo{pages}{1--14}.
\newblock


\bibitem[\protect\citeauthoryear{Jacobs, Johnson, and Mynatt}{Jacobs
  et~al\mbox{.}}{2018}]%
        {jacobs2018mypath}
\bibfield{author}{\bibinfo{person}{Maia Jacobs}, \bibinfo{person}{Jeremy
  Johnson}, {and} \bibinfo{person}{Elizabeth~D Mynatt}.}
  \bibinfo{year}{2018}\natexlab{}.
\newblock \showarticletitle{MyPath: Investigating breast cancer patients' use
  of personalized health information}.
\newblock \bibinfo{journal}{\emph{Proceedings of the ACM on Human-Computer
  Interaction}} \bibinfo{volume}{2}, \bibinfo{number}{CSCW}
  (\bibinfo{year}{2018}), \bibinfo{pages}{1--21}.
\newblock


\bibitem[\protect\citeauthoryear{Joseph}{Joseph}{2006}]%
        {joseph2006care}
\bibfield{author}{\bibinfo{person}{Amelia~M Joseph}.}
  \bibinfo{year}{2006}\natexlab{}.
\newblock \showarticletitle{Care coordination and telehealth technology in
  promoting self-management among chronically ill patients}.
\newblock \bibinfo{journal}{\emph{Telemedicine Journal \& e-Health}}
  \bibinfo{volume}{12}, \bibinfo{number}{2} (\bibinfo{year}{2006}),
  \bibinfo{pages}{156--159}.
\newblock


\bibitem[\protect\citeauthoryear{Kahlon, Yuan, Daigre, Meeks, Nelson,
  Piontkowski, Reuter, Sak, Turner, Weber, et~al\mbox{.}}{Kahlon
  et~al\mbox{.}}{2014}]%
        {kahlon2014use}
\bibfield{author}{\bibinfo{person}{Maninder Kahlon}, \bibinfo{person}{Leslie
  Yuan}, \bibinfo{person}{John Daigre}, \bibinfo{person}{Eric Meeks},
  \bibinfo{person}{Katie Nelson}, \bibinfo{person}{Cynthia Piontkowski},
  \bibinfo{person}{Katja Reuter}, \bibinfo{person}{Rachael Sak},
  \bibinfo{person}{Brian Turner}, \bibinfo{person}{Griffin~M Weber},
  {et~al\mbox{.}}} \bibinfo{year}{2014}\natexlab{}.
\newblock \showarticletitle{The use and significance of a research networking
  system}.
\newblock \bibinfo{journal}{\emph{Journal of Medical Internet Research}}
  \bibinfo{volume}{16}, \bibinfo{number}{2} (\bibinfo{year}{2014}),
  \bibinfo{pages}{e3137}.
\newblock


\bibitem[\protect\citeauthoryear{Karkar, Schroeder, Epstein, Pina, Scofield,
  Fogarty, Kientz, Munson, Vilardaga, and Zia}{Karkar et~al\mbox{.}}{2017}]%
        {karkar2017tummytrials}
\bibfield{author}{\bibinfo{person}{Ravi Karkar}, \bibinfo{person}{Jessica
  Schroeder}, \bibinfo{person}{Daniel~A Epstein}, \bibinfo{person}{Laura~R
  Pina}, \bibinfo{person}{Jeffrey Scofield}, \bibinfo{person}{James Fogarty},
  \bibinfo{person}{Julie~A Kientz}, \bibinfo{person}{Sean~A Munson},
  \bibinfo{person}{Roger Vilardaga}, {and} \bibinfo{person}{Jasmine Zia}.}
  \bibinfo{year}{2017}\natexlab{}.
\newblock \showarticletitle{Tummytrials: a feasibility study of using
  self-experimentation to detect individualized food triggers}. In
  \bibinfo{booktitle}{\emph{Proceedings of the 2017 CHI conference on human
  factors in computing systems}}. \bibinfo{pages}{6850--6863}.
\newblock


\bibitem[\protect\citeauthoryear{Kaziunas, Klinkman, and Ackerman}{Kaziunas
  et~al\mbox{.}}{2019}]%
        {kaziunas2019precarious}
\bibfield{author}{\bibinfo{person}{Elizabeth Kaziunas},
  \bibinfo{person}{Michael~S Klinkman}, {and} \bibinfo{person}{Mark~S
  Ackerman}.} \bibinfo{year}{2019}\natexlab{}.
\newblock \showarticletitle{Precarious interventions: Designing for ecologies
  of care}.
\newblock \bibinfo{journal}{\emph{Proceedings of the ACM on Human-Computer
  Interaction}} \bibinfo{volume}{3}, \bibinfo{number}{CSCW}
  (\bibinfo{year}{2019}), \bibinfo{pages}{1--27}.
\newblock


\bibitem[\protect\citeauthoryear{Kirchner, Schroeder, Fogarty, and
  Munson}{Kirchner et~al\mbox{.}}{2021}]%
        {kirchner2021they}
\bibfield{author}{\bibinfo{person}{Susanne Kirchner}, \bibinfo{person}{Jessica
  Schroeder}, \bibinfo{person}{James Fogarty}, {and} \bibinfo{person}{Sean~A
  Munson}.} \bibinfo{year}{2021}\natexlab{}.
\newblock \showarticletitle{“They don’t always think about that”:
  Translational Needs in the Design of Personal Health Informatics
  Applications}. In \bibinfo{booktitle}{\emph{Proceedings of the 2021 CHI
  Conference on Human Factors in Computing Systems}}. \bibinfo{pages}{1--16}.
\newblock


\bibitem[\protect\citeauthoryear{Klasnja, Consolvo, and Pratt}{Klasnja
  et~al\mbox{.}}{2011}]%
        {klasnja2011evaluate}
\bibfield{author}{\bibinfo{person}{Predrag Klasnja}, \bibinfo{person}{Sunny
  Consolvo}, {and} \bibinfo{person}{Wanda Pratt}.}
  \bibinfo{year}{2011}\natexlab{}.
\newblock \showarticletitle{How to evaluate technologies for health behavior
  change in HCI research}. In \bibinfo{booktitle}{\emph{Proceedings of the
  SIGCHI conference on human factors in computing systems}}.
  \bibinfo{pages}{3063--3072}.
\newblock


\bibitem[\protect\citeauthoryear{Kobb, Hoffman, Lodge, and Kline}{Kobb
  et~al\mbox{.}}{2003}]%
        {kobb2003enhancing}
\bibfield{author}{\bibinfo{person}{Rita Kobb}, \bibinfo{person}{Nannette
  Hoffman}, \bibinfo{person}{Robert Lodge}, {and} \bibinfo{person}{Sheri
  Kline}.} \bibinfo{year}{2003}\natexlab{}.
\newblock \showarticletitle{Enhancing elder chronic care through technology and
  care coordination: report from a pilot}.
\newblock \bibinfo{journal}{\emph{Telemedicine Journal and e-HEALTH}}
  \bibinfo{volume}{9}, \bibinfo{number}{2} (\bibinfo{year}{2003}),
  \bibinfo{pages}{189--195}.
\newblock


\bibitem[\protect\citeauthoryear{Kraut, Galegher, and Egido}{Kraut
  et~al\mbox{.}}{1987}]%
        {kraut1987relationships}
\bibfield{author}{\bibinfo{person}{Robert~E Kraut}, \bibinfo{person}{Jolene
  Galegher}, {and} \bibinfo{person}{Carmen Egido}.}
  \bibinfo{year}{1987}\natexlab{}.
\newblock \showarticletitle{Relationships and tasks in scientific research
  collaboration}.
\newblock \bibinfo{journal}{\emph{Human--Computer Interaction}}
  \bibinfo{volume}{3}, \bibinfo{number}{1} (\bibinfo{year}{1987}),
  \bibinfo{pages}{31--58}.
\newblock


\bibitem[\protect\citeauthoryear{Kruzan, Meyerhoff, Biernesser, Goldstein,
  Reddy, and Mohr}{Kruzan et~al\mbox{.}}{2021}]%
        {kruzan2021centering}
\bibfield{author}{\bibinfo{person}{Kaylee~Payne Kruzan}, \bibinfo{person}{Jonah
  Meyerhoff}, \bibinfo{person}{Candice Biernesser}, \bibinfo{person}{Tina
  Goldstein}, \bibinfo{person}{Madhu Reddy}, {and} \bibinfo{person}{David~C
  Mohr}.} \bibinfo{year}{2021}\natexlab{}.
\newblock \showarticletitle{Centering Lived Experience in Developing Digital
  Interventions for Suicide and Self-injurious Behaviors: User-Centered Design
  Approach}.
\newblock \bibinfo{journal}{\emph{JMIR mental health}} \bibinfo{volume}{8},
  \bibinfo{number}{12} (\bibinfo{year}{2021}), \bibinfo{pages}{e31367}.
\newblock


\bibitem[\protect\citeauthoryear{Kulp and Sarcevic}{Kulp and Sarcevic}{2018}]%
        {kulp2018design}
\bibfield{author}{\bibinfo{person}{Leah Kulp} {and} \bibinfo{person}{Aleksandra
  Sarcevic}.} \bibinfo{year}{2018}\natexlab{}.
\newblock \showarticletitle{Design in the “medical” wild: challenges of
  technology deployment}. In \bibinfo{booktitle}{\emph{Extended Abstracts of
  the 2018 CHI Conference on Human Factors in Computing Systems}}.
  \bibinfo{pages}{1--6}.
\newblock


\bibitem[\protect\citeauthoryear{Kulp, Sarcevic, Cheng, Zheng, and Burd}{Kulp
  et~al\mbox{.}}{2019}]%
        {kulp2019comparing}
\bibfield{author}{\bibinfo{person}{Leah Kulp}, \bibinfo{person}{Aleksandra
  Sarcevic}, \bibinfo{person}{Megan Cheng}, \bibinfo{person}{Yinan Zheng},
  {and} \bibinfo{person}{Randall~S Burd}.} \bibinfo{year}{2019}\natexlab{}.
\newblock \showarticletitle{Comparing the effects of paper and digital
  checklists on team performance in time-critical work}. In
  \bibinfo{booktitle}{\emph{Proceedings of the 2019 CHI Conference on Human
  Factors in Computing Systems}}. \bibinfo{pages}{1--13}.
\newblock


\bibitem[\protect\citeauthoryear{Kuo, Saran, Argentina, Heung, Bragg-Gresham,
  Chatoth, Gillespie, Krein, Wingard, Zheng, et~al\mbox{.}}{Kuo
  et~al\mbox{.}}{2019}]%
        {kuo2019development}
\bibfield{author}{\bibinfo{person}{Pei-Yi Kuo}, \bibinfo{person}{Rajiv Saran},
  \bibinfo{person}{Marissa Argentina}, \bibinfo{person}{Michael Heung},
  \bibinfo{person}{Jennifer~L Bragg-Gresham}, \bibinfo{person}{Dinesh Chatoth},
  \bibinfo{person}{Brenda Gillespie}, \bibinfo{person}{Sarah Krein},
  \bibinfo{person}{Rebecca Wingard}, \bibinfo{person}{Kai Zheng},
  {et~al\mbox{.}}} \bibinfo{year}{2019}\natexlab{}.
\newblock \showarticletitle{Development of a checklist for the prevention of
  intradialytic hypotension in hemodialysis care: Design considerations based
  on activity theory}. In \bibinfo{booktitle}{\emph{Proceedings of the 2019 CHI
  Conference on Human Factors in Computing Systems}}. \bibinfo{pages}{1--14}.
\newblock


\bibitem[\protect\citeauthoryear{Latulipe, Gatto, Nguyen, Miller, Quandt,
  Bertoni, Smith, and Arcury}{Latulipe et~al\mbox{.}}{2015}]%
        {latulipe2015design}
\bibfield{author}{\bibinfo{person}{Celine Latulipe}, \bibinfo{person}{Amy
  Gatto}, \bibinfo{person}{Ha~T Nguyen}, \bibinfo{person}{David~P Miller},
  \bibinfo{person}{Sara~A Quandt}, \bibinfo{person}{Alain~G Bertoni},
  \bibinfo{person}{Alden Smith}, {and} \bibinfo{person}{Thomas~A Arcury}.}
  \bibinfo{year}{2015}\natexlab{}.
\newblock \showarticletitle{Design considerations for patient portal adoption
  by low-income, older adults}. In \bibinfo{booktitle}{\emph{Proceedings of the
  33rd annual ACM conference on human factors in computing systems}}.
  \bibinfo{pages}{3859--3868}.
\newblock


\bibitem[\protect\citeauthoryear{Lazar, Su, Bardzell, and Bardzell}{Lazar
  et~al\mbox{.}}{2019}]%
        {lazar2019parting}
\bibfield{author}{\bibinfo{person}{Amanda Lazar},
  \bibinfo{person}{Norman~Makoto Su}, \bibinfo{person}{Jeffrey Bardzell}, {and}
  \bibinfo{person}{Shaowen Bardzell}.} \bibinfo{year}{2019}\natexlab{}.
\newblock \showarticletitle{Parting the Red Sea: sociotechnical systems and
  lived experiences of menopause}. In \bibinfo{booktitle}{\emph{Proceedings of
  the 2019 CHI conference on human factors in computing systems}}.
  \bibinfo{pages}{1--16}.
\newblock


\bibitem[\protect\citeauthoryear{Lee, Dourish, and Mark}{Lee
  et~al\mbox{.}}{2006}]%
        {lee2006human}
\bibfield{author}{\bibinfo{person}{Charlotte~P Lee}, \bibinfo{person}{Paul
  Dourish}, {and} \bibinfo{person}{Gloria Mark}.}
  \bibinfo{year}{2006}\natexlab{}.
\newblock \showarticletitle{The human infrastructure of cyberinfrastructure}.
  In \bibinfo{booktitle}{\emph{Proceedings of the 2006 20th anniversary
  conference on Computer supported cooperative work}}.
  \bibinfo{pages}{483--492}.
\newblock


\bibitem[\protect\citeauthoryear{Li, Liao, and Yen}{Li et~al\mbox{.}}{2013}]%
        {li2013co}
\bibfield{author}{\bibinfo{person}{Eldon~Y Li}, \bibinfo{person}{Chien~Hsiang
  Liao}, {and} \bibinfo{person}{Hsiuju~Rebecca Yen}.}
  \bibinfo{year}{2013}\natexlab{}.
\newblock \showarticletitle{Co-authorship networks and research impact: A
  social capital perspective}.
\newblock \bibinfo{journal}{\emph{Research Policy}} \bibinfo{volume}{42},
  \bibinfo{number}{9} (\bibinfo{year}{2013}), \bibinfo{pages}{1515--1530}.
\newblock


\bibitem[\protect\citeauthoryear{Liang, Munson, and Kientz}{Liang
  et~al\mbox{.}}{2021}]%
        {liang2021embracing}
\bibfield{author}{\bibinfo{person}{Calvin~A Liang}, \bibinfo{person}{Sean~A
  Munson}, {and} \bibinfo{person}{Julie~A Kientz}.}
  \bibinfo{year}{2021}\natexlab{}.
\newblock \showarticletitle{Embracing Four Tensions in Human-Computer
  Interaction Research with Marginalized People}.
\newblock \bibinfo{journal}{\emph{ACM Transactions on Computer-Human
  Interaction (TOCHI)}} \bibinfo{volume}{28}, \bibinfo{number}{2}
  (\bibinfo{year}{2021}), \bibinfo{pages}{1--47}.
\newblock


\bibitem[\protect\citeauthoryear{Lyon, Brewer, and Are{\'a}n}{Lyon
  et~al\mbox{.}}{2020}]%
        {lyon2020leveraging}
\bibfield{author}{\bibinfo{person}{Aaron~R Lyon}, \bibinfo{person}{Stephanie~K
  Brewer}, {and} \bibinfo{person}{Patricia~A Are{\'a}n}.}
  \bibinfo{year}{2020}\natexlab{}.
\newblock \showarticletitle{Leveraging human-centered design to implement
  modern psychological science: Return on an early investment.}
\newblock \bibinfo{journal}{\emph{American Psychologist}} \bibinfo{volume}{75},
  \bibinfo{number}{8} (\bibinfo{year}{2020}), \bibinfo{pages}{1067}.
\newblock


\bibitem[\protect\citeauthoryear{Lyon and Koerner}{Lyon and Koerner}{2016}]%
        {lyon2016user}
\bibfield{author}{\bibinfo{person}{Aaron~R Lyon} {and} \bibinfo{person}{Kelly
  Koerner}.} \bibinfo{year}{2016}\natexlab{}.
\newblock \showarticletitle{User-centered design for psychosocial intervention
  development and implementation.}
\newblock \bibinfo{journal}{\emph{Clinical Psychology: Science and Practice}}
  \bibinfo{volume}{23}, \bibinfo{number}{2} (\bibinfo{year}{2016}),
  \bibinfo{pages}{180}.
\newblock


\bibitem[\protect\citeauthoryear{MacLeod, Bastin, Liu, Siek, and
  Connelly}{MacLeod et~al\mbox{.}}{2017}]%
        {macleod2017grateful}
\bibfield{author}{\bibinfo{person}{Haley MacLeod}, \bibinfo{person}{Grace
  Bastin}, \bibinfo{person}{Leslie~S Liu}, \bibinfo{person}{Katie Siek}, {and}
  \bibinfo{person}{Kay Connelly}.} \bibinfo{year}{2017}\natexlab{}.
\newblock \showarticletitle{" Be Grateful You Don't Have a Real Disease"
  Understanding Rare Disease Relationships}. In
  \bibinfo{booktitle}{\emph{Proceedings of the 2017 CHI Conference on Human
  Factors in Computing Systems}}. \bibinfo{pages}{1660--1673}.
\newblock


\bibitem[\protect\citeauthoryear{Malinverni and Pares}{Malinverni and
  Pares}{2017}]%
        {malinverni2017autoethnographic}
\bibfield{author}{\bibinfo{person}{Laura Malinverni} {and}
  \bibinfo{person}{Narcis Pares}.} \bibinfo{year}{2017}\natexlab{}.
\newblock \showarticletitle{An autoethnographic approach to guide situated
  ethical decisions in participatory design with teenagers}.
\newblock \bibinfo{journal}{\emph{Interacting with Computers}}
  \bibinfo{volume}{29}, \bibinfo{number}{3} (\bibinfo{year}{2017}),
  \bibinfo{pages}{403--415}.
\newblock


\bibitem[\protect\citeauthoryear{Mamykina, M.~Smaldone, R.~Bakken, Elhadad,
  G.~Mitchell, M.~Desai, E.~Levine, N.~Tobin, Cassells, G.~Davidson,
  et~al\mbox{.}}{Mamykina et~al\mbox{.}}{2021}]%
        {mamykina2021scaling}
\bibfield{author}{\bibinfo{person}{Lena Mamykina}, \bibinfo{person}{Arlene
  M.~Smaldone}, \bibinfo{person}{Suzanne R.~Bakken}, \bibinfo{person}{Noemie
  Elhadad}, \bibinfo{person}{Elliot G.~Mitchell}, \bibinfo{person}{Pooja
  M.~Desai}, \bibinfo{person}{Matthew E.~Levine}, \bibinfo{person}{Jonathan
  N.~Tobin}, \bibinfo{person}{Andrea Cassells}, \bibinfo{person}{Patricia
  G.~Davidson}, {et~al\mbox{.}}} \bibinfo{year}{2021}\natexlab{}.
\newblock \showarticletitle{Scaling Up HCI Research: from Clinical Trials to
  Deployment in the Wild.}. In \bibinfo{booktitle}{\emph{Extended Abstracts of
  the 2021 CHI Conference on Human Factors in Computing Systems}}.
  \bibinfo{pages}{1--6}.
\newblock


\bibitem[\protect\citeauthoryear{Mao, Wang, Muller, Varshney, Baldini, Dugan,
  and Mojsilovi{\'c}}{Mao et~al\mbox{.}}{2019}]%
        {mao2019data}
\bibfield{author}{\bibinfo{person}{Yaoli Mao}, \bibinfo{person}{Dakuo Wang},
  \bibinfo{person}{Michael Muller}, \bibinfo{person}{Kush~R Varshney},
  \bibinfo{person}{Ioana Baldini}, \bibinfo{person}{Casey Dugan}, {and}
  \bibinfo{person}{Aleksandra Mojsilovi{\'c}}.}
  \bibinfo{year}{2019}\natexlab{}.
\newblock \showarticletitle{How data scientistswork together with domain
  experts in scientific collaborations: To find the right answer or to ask the
  right question?}
\newblock \bibinfo{journal}{\emph{Proceedings of the ACM on Human-Computer
  Interaction}} \bibinfo{volume}{3}, \bibinfo{number}{GROUP}
  (\bibinfo{year}{2019}), \bibinfo{pages}{1--23}.
\newblock


\bibitem[\protect\citeauthoryear{Maru{\v{s}}i{\'c}, Bo{\v{s}}njak, and
  Jeron{\v{c}}i{\'c}}{Maru{\v{s}}i{\'c} et~al\mbox{.}}{2011}]%
        {maruvsic2011systematic}
\bibfield{author}{\bibinfo{person}{Ana Maru{\v{s}}i{\'c}},
  \bibinfo{person}{Lana Bo{\v{s}}njak}, {and} \bibinfo{person}{Ana
  Jeron{\v{c}}i{\'c}}.} \bibinfo{year}{2011}\natexlab{}.
\newblock \showarticletitle{A systematic review of research on the meaning,
  ethics and practices of authorship across scholarly disciplines}.
\newblock \bibinfo{journal}{\emph{Plos one}} \bibinfo{volume}{6},
  \bibinfo{number}{9} (\bibinfo{year}{2011}), \bibinfo{pages}{e23477}.
\newblock


\bibitem[\protect\citeauthoryear{McHale, Ranwala, DiazGranados, Bagshaw,
  Schienke, and Blank}{McHale et~al\mbox{.}}{2019}]%
        {mchale2019promotion}
\bibfield{author}{\bibinfo{person}{Susan~M McHale},
  \bibinfo{person}{Damayanthi~Dayan Ranwala}, \bibinfo{person}{Deborah
  DiazGranados}, \bibinfo{person}{Dee Bagshaw}, \bibinfo{person}{Erich
  Schienke}, {and} \bibinfo{person}{Arthur~E Blank}.}
  \bibinfo{year}{2019}\natexlab{}.
\newblock \showarticletitle{Promotion and tenure policies for team science at
  colleges/schools of medicine}.
\newblock \bibinfo{journal}{\emph{Journal of Clinical and Translational
  Science}} \bibinfo{volume}{3}, \bibinfo{number}{5} (\bibinfo{year}{2019}),
  \bibinfo{pages}{245--252}.
\newblock


\bibitem[\protect\citeauthoryear{Merton}{Merton}{1968}]%
        {merton1968matthew}
\bibfield{author}{\bibinfo{person}{Robert~K Merton}.}
  \bibinfo{year}{1968}\natexlab{}.
\newblock \showarticletitle{The Matthew Effect in Science: The reward and
  communication systems of science are considered.}
\newblock \bibinfo{journal}{\emph{Science}} \bibinfo{volume}{159},
  \bibinfo{number}{3810} (\bibinfo{year}{1968}), \bibinfo{pages}{56--63}.
\newblock


\bibitem[\protect\citeauthoryear{Merton}{Merton}{1988}]%
        {merton1988matthew}
\bibfield{author}{\bibinfo{person}{Robert~K Merton}.}
  \bibinfo{year}{1988}\natexlab{}.
\newblock \showarticletitle{The Matthew effect in science, II: Cumulative
  advantage and the symbolism of intellectual property}.
\newblock \bibinfo{journal}{\emph{isis}} \bibinfo{volume}{79},
  \bibinfo{number}{4} (\bibinfo{year}{1988}), \bibinfo{pages}{606--623}.
\newblock


\bibitem[\protect\citeauthoryear{Mishra, Klasnja, MacDuffie~Woodburn, Hekler,
  Omberg, Kellen, and Mangravite}{Mishra et~al\mbox{.}}{2019}]%
        {mishra2019supporting}
\bibfield{author}{\bibinfo{person}{Sonali~R Mishra}, \bibinfo{person}{Predrag
  Klasnja}, \bibinfo{person}{John MacDuffie~Woodburn}, \bibinfo{person}{Eric~B
  Hekler}, \bibinfo{person}{Larsson Omberg}, \bibinfo{person}{Michael Kellen},
  {and} \bibinfo{person}{Lara Mangravite}.} \bibinfo{year}{2019}\natexlab{}.
\newblock \showarticletitle{Supporting coping with parkinson's disease through
  self tracking}. In \bibinfo{booktitle}{\emph{Proceedings of the 2019 CHI
  Conference on Human Factors in Computing Systems}}. \bibinfo{pages}{1--16}.
\newblock


\bibitem[\protect\citeauthoryear{Mitchell and Mamykina}{Mitchell and
  Mamykina}{2021}]%
        {mitchell2021curtain}
\bibfield{author}{\bibinfo{person}{Elliot Mitchell} {and} \bibinfo{person}{Lena
  Mamykina}.} \bibinfo{year}{2021}\natexlab{}.
\newblock \showarticletitle{From the Curtain to Kansas: Conducting Wizard-of-Oz
  Studies in the Wild}. In \bibinfo{booktitle}{\emph{Extended Abstracts of the
  2021 CHI Conference on Human Factors in Computing Systems}}.
  \bibinfo{pages}{1--6}.
\newblock


\bibitem[\protect\citeauthoryear{Mohr, Burns, Schueller, Clarke, and
  Klinkman}{Mohr et~al\mbox{.}}{2013}]%
        {mohr2013behavioral}
\bibfield{author}{\bibinfo{person}{David~C Mohr},
  \bibinfo{person}{Michelle~Nicole Burns}, \bibinfo{person}{Stephen~M
  Schueller}, \bibinfo{person}{Gregory Clarke}, {and} \bibinfo{person}{Michael
  Klinkman}.} \bibinfo{year}{2013}\natexlab{}.
\newblock \showarticletitle{Behavioral intervention technologies: evidence
  review and recommendations for future research in mental health}.
\newblock \bibinfo{journal}{\emph{General hospital psychiatry}}
  \bibinfo{volume}{35}, \bibinfo{number}{4} (\bibinfo{year}{2013}),
  \bibinfo{pages}{332--338}.
\newblock


\bibitem[\protect\citeauthoryear{Moody, Slocumb, Berg, and Jackson}{Moody
  et~al\mbox{.}}{2004}]%
        {moody2004electronic}
\bibfield{author}{\bibinfo{person}{Linda~E Moody}, \bibinfo{person}{Elaine
  Slocumb}, \bibinfo{person}{Bruce Berg}, {and} \bibinfo{person}{Donna
  Jackson}.} \bibinfo{year}{2004}\natexlab{}.
\newblock \showarticletitle{Electronic health records documentation in nursing:
  nurses' perceptions, attitudes, and preferences}.
\newblock \bibinfo{journal}{\emph{CIN: Computers, Informatics, Nursing}}
  \bibinfo{volume}{22}, \bibinfo{number}{6} (\bibinfo{year}{2004}),
  \bibinfo{pages}{337--344}.
\newblock


\bibitem[\protect\citeauthoryear{Nadal, McCully, Doherty, Sas, and
  Doherty}{Nadal et~al\mbox{.}}{2022}]%
        {nadal2022tac}
\bibfield{author}{\bibinfo{person}{Camille Nadal}, \bibinfo{person}{Shane
  McCully}, \bibinfo{person}{Kevin Doherty}, \bibinfo{person}{Corina Sas},
  {and} \bibinfo{person}{Gavin Doherty}.} \bibinfo{year}{2022}\natexlab{}.
\newblock \showarticletitle{The TAC toolkit: supporting the design for user
  acceptance of health technologies from a macro-temporal perspective}. In
  \bibinfo{booktitle}{\emph{Proceedings of the CHI Conference on Human Factors
  in Computing Systems}}.
\newblock


\bibitem[\protect\citeauthoryear{Nagel, Koch, Guimond, Glavin, and
  Geller}{Nagel et~al\mbox{.}}{2013}]%
        {nagel2013building}
\bibfield{author}{\bibinfo{person}{Joan~D Nagel}, \bibinfo{person}{Abby Koch},
  \bibinfo{person}{Jennifer~M Guimond}, \bibinfo{person}{Sarah Glavin}, {and}
  \bibinfo{person}{Stacie Geller}.} \bibinfo{year}{2013}\natexlab{}.
\newblock \showarticletitle{Building the women's health research workforce:
  fostering interdisciplinary research approaches in women's health}.
\newblock \bibinfo{journal}{\emph{Global Advances in Health and Medicine}}
  \bibinfo{volume}{2}, \bibinfo{number}{5} (\bibinfo{year}{2013}),
  \bibinfo{pages}{24--29}.
\newblock


\bibitem[\protect\citeauthoryear{Nakikj and Mamykina}{Nakikj and
  Mamykina}{2017}]%
        {nakikj2017park}
\bibfield{author}{\bibinfo{person}{Drashko Nakikj} {and} \bibinfo{person}{Lena
  Mamykina}.} \bibinfo{year}{2017}\natexlab{}.
\newblock \showarticletitle{A park or a highway: Overcoming tensions in
  designing for socio-emotional and informational needs in online health
  communities}. In \bibinfo{booktitle}{\emph{Proceedings of the 2017 ACM
  Conference on Computer Supported Cooperative Work and Social Computing}}.
  \bibinfo{pages}{1304--1319}.
\newblock


\bibitem[\protect\citeauthoryear{Ogbonnaya-Ogburu, Smith, To, and
  Toyama}{Ogbonnaya-Ogburu et~al\mbox{.}}{2020}]%
        {ogbonnaya2020critical}
\bibfield{author}{\bibinfo{person}{Ihudiya~Finda Ogbonnaya-Ogburu},
  \bibinfo{person}{Angela~DR Smith}, \bibinfo{person}{Alexandra To}, {and}
  \bibinfo{person}{Kentaro Toyama}.} \bibinfo{year}{2020}\natexlab{}.
\newblock \showarticletitle{Critical race theory for HCI}. In
  \bibinfo{booktitle}{\emph{Proceedings of the 2020 CHI conference on human
  factors in computing systems}}. \bibinfo{pages}{1--16}.
\newblock


\bibitem[\protect\citeauthoryear{O'Leary, Schueller, Wobbrock, and
  Pratt}{O'Leary et~al\mbox{.}}{2018}]%
        {o2018suddenly}
\bibfield{author}{\bibinfo{person}{Kathleen O'Leary},
  \bibinfo{person}{Stephen~M Schueller}, \bibinfo{person}{Jacob~O Wobbrock},
  {and} \bibinfo{person}{Wanda Pratt}.} \bibinfo{year}{2018}\natexlab{}.
\newblock \showarticletitle{“Suddenly, we got to become therapists for each
  other” Designing Peer Support Chats for Mental Health}. In
  \bibinfo{booktitle}{\emph{Proceedings of the 2018 CHI Conference on Human
  Factors in Computing Systems}}. \bibinfo{pages}{1--14}.
\newblock


\bibitem[\protect\citeauthoryear{Olson and Olson}{Olson and Olson}{2000}]%
        {olson2000distance}
\bibfield{author}{\bibinfo{person}{Gary~M Olson} {and}
  \bibinfo{person}{Judith~S Olson}.} \bibinfo{year}{2000}\natexlab{}.
\newblock \showarticletitle{Distance matters}.
\newblock \bibinfo{journal}{\emph{Human--computer interaction}}
  \bibinfo{volume}{15}, \bibinfo{number}{2-3} (\bibinfo{year}{2000}),
  \bibinfo{pages}{139--178}.
\newblock


\bibitem[\protect\citeauthoryear{Olson and Olson}{Olson and Olson}{2013}]%
        {olson2013working}
\bibfield{author}{\bibinfo{person}{Judith~S Olson} {and}
  \bibinfo{person}{Gary~M Olson}.} \bibinfo{year}{2013}\natexlab{}.
\newblock \showarticletitle{Working together apart: Collaboration over the
  internet}.
\newblock \bibinfo{journal}{\emph{Synthesis Lectures on Human-Centered
  Informatics}} \bibinfo{volume}{6}, \bibinfo{number}{5}
  (\bibinfo{year}{2013}), \bibinfo{pages}{1--151}.
\newblock


\bibitem[\protect\citeauthoryear{on~Science, Policy, of~Medicine~(US),
  on~Facilitating Interdisciplinary~Research, of~Engineering~(US), and
  of~Sciences~(US)}{on~Science et~al\mbox{.}}{2004}]%
        {engineering2004facilitating}
\bibfield{author}{\bibinfo{person}{Engineering~Committee on Science},
  \bibinfo{person}{Public Policy}, \bibinfo{person}{Institute of
  Medicine~(US)}, \bibinfo{person}{National Academies (US).~Committee on
  Facilitating Interdisciplinary~Research}, \bibinfo{person}{National~Academy
  of Engineering~(US)}, {and} \bibinfo{person}{National~Academy of
  Sciences~(US)}.} \bibinfo{year}{2004}\natexlab{}.
\newblock \bibinfo{booktitle}{\emph{Facilitating interdisciplinary research}}.
\newblock \bibinfo{publisher}{National Academies Press}.
\newblock


\bibitem[\protect\citeauthoryear{Pater, Coupe, Pfafman, Phelan, Toscos, and
  Jacobs}{Pater et~al\mbox{.}}{2021}]%
        {pater2021standardizing}
\bibfield{author}{\bibinfo{person}{Jessica Pater}, \bibinfo{person}{Amanda
  Coupe}, \bibinfo{person}{Rachel Pfafman}, \bibinfo{person}{Chanda Phelan},
  \bibinfo{person}{Tammy Toscos}, {and} \bibinfo{person}{Maia Jacobs}.}
  \bibinfo{year}{2021}\natexlab{}.
\newblock \showarticletitle{Standardizing Reporting of Participant Compensation
  in HCI: A Systematic Literature Review and Recommendations for the Field}. In
  \bibinfo{booktitle}{\emph{Proceedings of the 2021 CHI Conference on Human
  Factors in Computing Systems}}. \bibinfo{pages}{1--16}.
\newblock


\bibitem[\protect\citeauthoryear{Pollack, Backonja, Miller, Mishra, Khelifi,
  Kendall, and Pratt}{Pollack et~al\mbox{.}}{2016}]%
        {pollack2016closing}
\bibfield{author}{\bibinfo{person}{Ari~H Pollack}, \bibinfo{person}{Uba
  Backonja}, \bibinfo{person}{Andrew~D Miller}, \bibinfo{person}{Sonali~R
  Mishra}, \bibinfo{person}{Maher Khelifi}, \bibinfo{person}{Logan Kendall},
  {and} \bibinfo{person}{Wanda Pratt}.} \bibinfo{year}{2016}\natexlab{}.
\newblock \showarticletitle{Closing the gap: supporting patients' transition to
  self-management after hospitalization}. In
  \bibinfo{booktitle}{\emph{Proceedings of the 2016 CHI Conference on Human
  Factors in Computing Systems}}. \bibinfo{pages}{5324--5336}.
\newblock


\bibitem[\protect\citeauthoryear{Repko and Szostak}{Repko and Szostak}{2020}]%
        {repko2020interdisciplinary}
\bibfield{author}{\bibinfo{person}{Allen~F Repko} {and} \bibinfo{person}{Rick
  Szostak}.} \bibinfo{year}{2020}\natexlab{}.
\newblock \bibinfo{booktitle}{\emph{Interdisciplinary research: Process and
  theory}}.
\newblock \bibinfo{publisher}{Sage Publications}.
\newblock


\bibitem[\protect\citeauthoryear{Rolland, Paine, and Lee}{Rolland
  et~al\mbox{.}}{2014}]%
        {rolland2014work}
\bibfield{author}{\bibinfo{person}{Betsy Rolland}, \bibinfo{person}{Drew
  Paine}, {and} \bibinfo{person}{Charlotte~P Lee}.}
  \bibinfo{year}{2014}\natexlab{}.
\newblock \showarticletitle{Work practices in coordinating center enabled
  networks (CCENs)}. In \bibinfo{booktitle}{\emph{Proceedings of the 18th
  International Conference on Supporting Group Work}}.
  \bibinfo{pages}{194--203}.
\newblock


\bibitem[\protect\citeauthoryear{Rothschild, Dietrich, Ball, Wurtz,
  Farish-Hunt, and Cortes-Comerer}{Rothschild et~al\mbox{.}}{2005}]%
        {rothschild2005leveraging}
\bibfield{author}{\bibinfo{person}{Adam~S Rothschild}, \bibinfo{person}{Linda
  Dietrich}, \bibinfo{person}{Marion~J Ball}, \bibinfo{person}{Heidi Wurtz},
  \bibinfo{person}{Holly Farish-Hunt}, {and} \bibinfo{person}{Nhora
  Cortes-Comerer}.} \bibinfo{year}{2005}\natexlab{}.
\newblock \showarticletitle{Leveraging systems thinking to design
  patient-centered clinical documentation systems}.
\newblock \bibinfo{journal}{\emph{International Journal of Medical
  Informatics}} \bibinfo{volume}{74}, \bibinfo{number}{5}
  (\bibinfo{year}{2005}), \bibinfo{pages}{395--398}.
\newblock


\bibitem[\protect\citeauthoryear{Saksono, Castaneda-Sceppa, Hoffman, Seif
  El-Nasr, and Parker}{Saksono et~al\mbox{.}}{2021}]%
        {saksono2021storymap}
\bibfield{author}{\bibinfo{person}{Herman Saksono}, \bibinfo{person}{Carmen
  Castaneda-Sceppa}, \bibinfo{person}{Jessica~A Hoffman}, \bibinfo{person}{Magy
  Seif El-Nasr}, {and} \bibinfo{person}{Andrea Parker}.}
  \bibinfo{year}{2021}\natexlab{}.
\newblock \showarticletitle{StoryMap: Using Social Modeling and Self-Modeling
  to Support Physical Activity Among Families of Low-SES Backgrounds}. In
  \bibinfo{booktitle}{\emph{Proceedings of the 2021 CHI Conference on Human
  Factors in Computing Systems}}. \bibinfo{pages}{1--14}.
\newblock


\bibitem[\protect\citeauthoryear{Salazar, Lant, Fiore, and Salas}{Salazar
  et~al\mbox{.}}{2012}]%
        {salazar2012facilitating}
\bibfield{author}{\bibinfo{person}{Maritza~R Salazar},
  \bibinfo{person}{Theresa~K Lant}, \bibinfo{person}{Stephen~M Fiore}, {and}
  \bibinfo{person}{Eduardo Salas}.} \bibinfo{year}{2012}\natexlab{}.
\newblock \showarticletitle{Facilitating innovation in diverse science teams
  through integrative capacity}.
\newblock \bibinfo{journal}{\emph{Small Group Research}} \bibinfo{volume}{43},
  \bibinfo{number}{5} (\bibinfo{year}{2012}), \bibinfo{pages}{527--558}.
\newblock


\bibitem[\protect\citeauthoryear{Salda{\~n}a}{Salda{\~n}a}{2021}]%
        {saldana2021coding}
\bibfield{author}{\bibinfo{person}{Johnny Salda{\~n}a}.}
  \bibinfo{year}{2021}\natexlab{}.
\newblock \bibinfo{booktitle}{\emph{The coding manual for qualitative
  researchers}}.
\newblock \bibinfo{publisher}{sage}.
\newblock


\bibitem[\protect\citeauthoryear{Schroeder, Chung, Epstein, Karkar, Parsons,
  Murinova, Fogarty, and Munson}{Schroeder et~al\mbox{.}}{2018}]%
        {schroeder2018examining}
\bibfield{author}{\bibinfo{person}{Jessica Schroeder},
  \bibinfo{person}{Chia-Fang Chung}, \bibinfo{person}{Daniel~A Epstein},
  \bibinfo{person}{Ravi Karkar}, \bibinfo{person}{Adele Parsons},
  \bibinfo{person}{Natalia Murinova}, \bibinfo{person}{James Fogarty}, {and}
  \bibinfo{person}{Sean~A Munson}.} \bibinfo{year}{2018}\natexlab{}.
\newblock \showarticletitle{Examining self-tracking by people with migraine:
  goals, needs, and opportunities in a chronic health condition}. In
  \bibinfo{booktitle}{\emph{Proceedings of the 2018 Designing Interactive
  Systems Conference}}. \bibinfo{pages}{135--148}.
\newblock


\bibitem[\protect\citeauthoryear{Seo, Berry, Bhagane, Choi, Buyuktur, and
  Park}{Seo et~al\mbox{.}}{2019}]%
        {seo2019balancing}
\bibfield{author}{\bibinfo{person}{Woosuk Seo}, \bibinfo{person}{Andrew~BL
  Berry}, \bibinfo{person}{Prachi Bhagane}, \bibinfo{person}{Sung~Won Choi},
  \bibinfo{person}{Ayse~G Buyuktur}, {and} \bibinfo{person}{Sun~Young Park}.}
  \bibinfo{year}{2019}\natexlab{}.
\newblock \showarticletitle{Balancing tensions between caregiving and parenting
  responsibilities in pediatric patient care}.
\newblock \bibinfo{journal}{\emph{Proceedings of the ACM on Human-Computer
  Interaction}} \bibinfo{volume}{3}, \bibinfo{number}{CSCW}
  (\bibinfo{year}{2019}), \bibinfo{pages}{1--24}.
\newblock


\bibitem[\protect\citeauthoryear{Shneiderman}{Shneiderman}{2016}]%
        {shneiderman2016teamwork}
\bibfield{author}{\bibinfo{person}{Ben Shneiderman}.}
  \bibinfo{year}{2016}\natexlab{}.
\newblock \showarticletitle{Teamwork in computing research}.
\newblock \bibinfo{journal}{\emph{Commun. ACM}} \bibinfo{volume}{59},
  \bibinfo{number}{8} (\bibinfo{year}{2016}), \bibinfo{pages}{30--31}.
\newblock


\bibitem[\protect\citeauthoryear{Shulman}{Shulman}{2001}]%
        {shulman2001carnegie}
\bibfield{author}{\bibinfo{person}{Lee~S Shulman}.}
  \bibinfo{year}{2001}\natexlab{}.
\newblock \showarticletitle{The Carnegie classification of institutions of
  higher education}.
\newblock \bibinfo{journal}{\emph{Menlo Park: Carnegie Publication}}
  (\bibinfo{year}{2001}).
\newblock


\bibitem[\protect\citeauthoryear{Singh, Newhouse, Gibbs, Blandford, Chen,
  Briggs, Mentis, Sellen, and Bardram}{Singh et~al\mbox{.}}{2017}]%
        {singh2017hci}
\bibfield{author}{\bibinfo{person}{Aneesha Singh}, \bibinfo{person}{Nikki
  Newhouse}, \bibinfo{person}{Jo Gibbs}, \bibinfo{person}{Ann~E Blandford},
  \bibinfo{person}{Yunan Chen}, \bibinfo{person}{Pam Briggs},
  \bibinfo{person}{Helena Mentis}, \bibinfo{person}{Kate~M Sellen}, {and}
  \bibinfo{person}{Jakob~E Bardram}.} \bibinfo{year}{2017}\natexlab{}.
\newblock \showarticletitle{HCI and health: Learning from interdisciplinary
  interactions}. In \bibinfo{booktitle}{\emph{Proceedings of the 2017 CHI
  Conference Extended Abstracts on Human Factors in Computing Systems}}.
  \bibinfo{pages}{1322--1325}.
\newblock


\bibitem[\protect\citeauthoryear{Smith, Nevarez, and Zhu}{Smith
  et~al\mbox{.}}{2020}]%
        {smith2020disseminating}
\bibfield{author}{\bibinfo{person}{C~Estelle Smith}, \bibinfo{person}{Eduardo
  Nevarez}, {and} \bibinfo{person}{Haiyi Zhu}.}
  \bibinfo{year}{2020}\natexlab{}.
\newblock \showarticletitle{Disseminating Research News in HCI: Perceived
  Hazards, How-To's, and Opportunities for Innovation}. In
  \bibinfo{booktitle}{\emph{Proceedings of the 2020 CHI Conference on Human
  Factors in Computing Systems}}. \bibinfo{pages}{1--13}.
\newblock


\bibitem[\protect\citeauthoryear{Smith, Wang, Karumur, and Zhu}{Smith
  et~al\mbox{.}}{2018}]%
        {smith2018breaking}
\bibfield{author}{\bibinfo{person}{C~Estelle Smith}, \bibinfo{person}{Xinyi
  Wang}, \bibinfo{person}{Raghav~Pavan Karumur}, {and} \bibinfo{person}{Haiyi
  Zhu}.} \bibinfo{year}{2018}\natexlab{}.
\newblock \showarticletitle{[Un] breaking News: Design Opportunities for
  Enhancing Collaboration in Scientific Media Production}. In
  \bibinfo{booktitle}{\emph{Proceedings of the 2018 CHI Conference on Human
  Factors in Computing Systems}}. \bibinfo{pages}{1--13}.
\newblock


\bibitem[\protect\citeauthoryear{Spring, Hall, Moller, and
  Falk-Krzesinski}{Spring et~al\mbox{.}}{2012}]%
        {spring2012emerging}
\bibfield{author}{\bibinfo{person}{Bonnie Spring}, \bibinfo{person}{Kara~L
  Hall}, \bibinfo{person}{Arlen~C Moller}, {and} \bibinfo{person}{Holly~J
  Falk-Krzesinski}.} \bibinfo{year}{2012}\natexlab{}.
\newblock \bibinfo{title}{An emerging science and praxis for research and
  practice teams}.
\newblock
\newblock


\bibitem[\protect\citeauthoryear{Spring, Pfammatter, and Conroy}{Spring
  et~al\mbox{.}}{2019}]%
        {spring2019continuing}
\bibfield{author}{\bibinfo{person}{Bonnie~J Spring},
  \bibinfo{person}{Angela~Fidler Pfammatter}, {and} \bibinfo{person}{David~E
  Conroy}.} \bibinfo{year}{2019}\natexlab{}.
\newblock \showarticletitle{Continuing professional development for team
  science}.
\newblock In \bibinfo{booktitle}{\emph{Strategies for Team Science Success}}.
  \bibinfo{publisher}{Springer}, \bibinfo{pages}{445--453}.
\newblock


\bibitem[\protect\citeauthoryear{Stokols, Hall, Taylor, and Moser}{Stokols
  et~al\mbox{.}}{2008}]%
        {stokols2008science}
\bibfield{author}{\bibinfo{person}{Daniel Stokols}, \bibinfo{person}{Kara~L
  Hall}, \bibinfo{person}{Brandie~K Taylor}, {and} \bibinfo{person}{Richard~P
  Moser}.} \bibinfo{year}{2008}\natexlab{}.
\newblock \showarticletitle{The science of team science: overview of the field
  and introduction to the supplement}.
\newblock \bibinfo{journal}{\emph{American journal of preventive medicine}}
  \bibinfo{volume}{35}, \bibinfo{number}{2} (\bibinfo{year}{2008}),
  \bibinfo{pages}{S77--S89}.
\newblock


\bibitem[\protect\citeauthoryear{Stokols, Hall, and Vogel}{Stokols
  et~al\mbox{.}}{2013}]%
        {stokols2013transdisciplinary}
\bibfield{author}{\bibinfo{person}{Daniel Stokols}, \bibinfo{person}{Kara~L
  Hall}, {and} \bibinfo{person}{Amanda~L Vogel}.}
  \bibinfo{year}{2013}\natexlab{}.
\newblock \showarticletitle{Transdisciplinary public health: definitions, core
  characteristics, and strategies for success}.
\newblock \bibinfo{journal}{\emph{Transdisciplinary public health: research,
  methods, and practice. San Francisco: Jossey-Bass}} (\bibinfo{year}{2013}),
  \bibinfo{pages}{3--30}.
\newblock


\bibitem[\protect\citeauthoryear{Toubia}{Toubia}{2006}]%
        {toubia2006idea}
\bibfield{author}{\bibinfo{person}{Olivier Toubia}.}
  \bibinfo{year}{2006}\natexlab{}.
\newblock \showarticletitle{Idea generation, creativity, and incentives}.
\newblock \bibinfo{journal}{\emph{Marketing science}} \bibinfo{volume}{25},
  \bibinfo{number}{5} (\bibinfo{year}{2006}), \bibinfo{pages}{411--425}.
\newblock


\bibitem[\protect\citeauthoryear{Tscharntke, Hochberg, Rand, Resh, and
  Krauss}{Tscharntke et~al\mbox{.}}{2007}]%
        {tscharntke2007author}
\bibfield{author}{\bibinfo{person}{Teja Tscharntke}, \bibinfo{person}{Michael~E
  Hochberg}, \bibinfo{person}{Tatyana~A Rand}, \bibinfo{person}{Vincent~H
  Resh}, {and} \bibinfo{person}{Jochen Krauss}.}
  \bibinfo{year}{2007}\natexlab{}.
\newblock \showarticletitle{Author sequence and credit for contributions in
  multiauthored publications}.
\newblock \bibinfo{journal}{\emph{PLoS biology}} \bibinfo{volume}{5},
  \bibinfo{number}{1} (\bibinfo{year}{2007}), \bibinfo{pages}{e18}.
\newblock


\bibitem[\protect\citeauthoryear{Veinot, Ancker, Cole-Lewis, Mynatt, Parker,
  Siek, and Mamykina}{Veinot et~al\mbox{.}}{2019}]%
        {veinot2019leveling}
\bibfield{author}{\bibinfo{person}{Tiffany~C Veinot},
  \bibinfo{person}{Jessica~S Ancker}, \bibinfo{person}{Heather Cole-Lewis},
  \bibinfo{person}{Elizabeth~D Mynatt}, \bibinfo{person}{Andrea~G Parker},
  \bibinfo{person}{Katie~A Siek}, {and} \bibinfo{person}{Lena Mamykina}.}
  \bibinfo{year}{2019}\natexlab{}.
\newblock \showarticletitle{Leveling up: on the potential of upstream health
  informatics interventions to enhance health equity}.
\newblock \bibinfo{journal}{\emph{Medical care}}  \bibinfo{volume}{57}
  (\bibinfo{year}{2019}), \bibinfo{pages}{S108--S114}.
\newblock


\bibitem[\protect\citeauthoryear{Verdezoto, Carpio-Arias, Carpio-Arias,
  Mackintosh, Eslambolchilar, Delgado, Andrade, and V{\'a}sconez}{Verdezoto
  et~al\mbox{.}}{2020}]%
        {verdezoto2020indigenous}
\bibfield{author}{\bibinfo{person}{Nervo Verdezoto}, \bibinfo{person}{Francisca
  Carpio-Arias}, \bibinfo{person}{Valeria Carpio-Arias},
  \bibinfo{person}{Nicola Mackintosh}, \bibinfo{person}{Parisa Eslambolchilar},
  \bibinfo{person}{Ver{\'o}nica Delgado}, \bibinfo{person}{Catherine Andrade},
  {and} \bibinfo{person}{Galo V{\'a}sconez}.} \bibinfo{year}{2020}\natexlab{}.
\newblock \showarticletitle{Indigenous Women Managing Pregnancy Complications
  in Rural Ecuador: Barriers and Opportunities to Enhance Antenatal Care}. In
  \bibinfo{booktitle}{\emph{Proceedings of the 11th Nordic Conference on
  Human-Computer Interaction: Shaping Experiences, Shaping Society}}.
  \bibinfo{pages}{1--9}.
\newblock


\bibitem[\protect\citeauthoryear{Vitak, Shilton, and Ashktorab}{Vitak
  et~al\mbox{.}}{2016}]%
        {vitak2016beyond}
\bibfield{author}{\bibinfo{person}{Jessica Vitak}, \bibinfo{person}{Katie
  Shilton}, {and} \bibinfo{person}{Zahra Ashktorab}.}
  \bibinfo{year}{2016}\natexlab{}.
\newblock \showarticletitle{Beyond the Belmont principles: Ethical challenges,
  practices, and beliefs in the online data research community}. In
  \bibinfo{booktitle}{\emph{Proceedings of the 19th ACM conference on
  computer-supported cooperative work \& social computing}}.
  \bibinfo{pages}{941--953}.
\newblock


\bibitem[\protect\citeauthoryear{Vogel, Feng, Oh, Hall, Stipelman, Stokols,
  Okamoto, Perna, Moser, and Nebeling}{Vogel et~al\mbox{.}}{2012}]%
        {vogel2012influence}
\bibfield{author}{\bibinfo{person}{Amanda~L Vogel}, \bibinfo{person}{Annie
  Feng}, \bibinfo{person}{April Oh}, \bibinfo{person}{Kara~L Hall},
  \bibinfo{person}{Brooke~A Stipelman}, \bibinfo{person}{Daniel Stokols},
  \bibinfo{person}{Janet Okamoto}, \bibinfo{person}{Frank~M Perna},
  \bibinfo{person}{Richard Moser}, {and} \bibinfo{person}{Linda Nebeling}.}
  \bibinfo{year}{2012}\natexlab{}.
\newblock \showarticletitle{Influence of a National Cancer Institute
  transdisciplinary research and training initiative on trainees'
  transdisciplinary research competencies and scholarly productivity}.
\newblock \bibinfo{journal}{\emph{Translational behavioral medicine}}
  \bibinfo{volume}{2}, \bibinfo{number}{4} (\bibinfo{year}{2012}),
  \bibinfo{pages}{459--468}.
\newblock


\bibitem[\protect\citeauthoryear{Walling and Walling}{Walling and
  Walling}{2018}]%
        {walling2018academic}
\bibfield{author}{\bibinfo{person}{Anne Walling} {and}
  \bibinfo{person}{Walling}.} \bibinfo{year}{2018}\natexlab{}.
\newblock \bibinfo{booktitle}{\emph{Academic promotion for clinicians}}.
\newblock \bibinfo{publisher}{Springer}.
\newblock


\bibitem[\protect\citeauthoryear{Weibel, Unertl, and Boll}{Weibel
  et~al\mbox{.}}{2019}]%
        {weibel2019symposium}
\bibfield{author}{\bibinfo{person}{Nadir Weibel}, \bibinfo{person}{Kim Unertl},
  {and} \bibinfo{person}{Susanne Boll}.} \bibinfo{year}{2019}\natexlab{}.
\newblock \showarticletitle{Symposium: WISH-Workgroup on Interactive Systems in
  Healthcare}. In \bibinfo{booktitle}{\emph{Extended Abstracts of the 2019 CHI
  Conference on Human Factors in Computing Systems}}. \bibinfo{pages}{1--8}.
\newblock


\bibitem[\protect\citeauthoryear{Williams, Hayes, Guo, Rahmani, and
  Dutt}{Williams et~al\mbox{.}}{2020}]%
        {williams2020hci}
\bibfield{author}{\bibinfo{person}{Lucretia Williams},
  \bibinfo{person}{Gillian~R Hayes}, \bibinfo{person}{Yuqing Guo},
  \bibinfo{person}{Amir Rahmani}, {and} \bibinfo{person}{Nikil Dutt}.}
  \bibinfo{year}{2020}\natexlab{}.
\newblock \showarticletitle{HCI and mHealth wearable tech: A multidisciplinary
  research challenge}. In \bibinfo{booktitle}{\emph{Extended Abstracts of the
  2020 CHI Conference on Human Factors in Computing Systems}}.
  \bibinfo{pages}{1--7}.
\newblock


\bibitem[\protect\citeauthoryear{Winner}{Winner}{1980}]%
        {winner1980artifacts}
\bibfield{author}{\bibinfo{person}{Langdon Winner}.}
  \bibinfo{year}{1980}\natexlab{}.
\newblock \showarticletitle{Do artifacts have politics?}
\newblock \bibinfo{journal}{\emph{Daedalus}} (\bibinfo{year}{1980}),
  \bibinfo{pages}{121--136}.
\newblock


\bibitem[\protect\citeauthoryear{Zerhouni et~al\mbox{.}}{Zerhouni
  et~al\mbox{.}}{2005}]%
        {zerhouni2005translational}
\bibfield{author}{\bibinfo{person}{Elias~A Zerhouni} {et~al\mbox{.}}}
  \bibinfo{year}{2005}\natexlab{}.
\newblock \showarticletitle{Translational and clinical science-time for a new
  vision}.
\newblock \bibinfo{journal}{\emph{New England Journal of Medicine}}
  \bibinfo{volume}{353}, \bibinfo{number}{15} (\bibinfo{year}{2005}),
  \bibinfo{pages}{1621}.
\newblock


\bibitem[\protect\citeauthoryear{Zhang, Sun, Padilla, Barua, Bertini, and
  Parker}{Zhang et~al\mbox{.}}{2021}]%
        {zhang2021mapping}
\bibfield{author}{\bibinfo{person}{Yixuan Zhang}, \bibinfo{person}{Yifan Sun},
  \bibinfo{person}{Lace Padilla}, \bibinfo{person}{Sumit Barua},
  \bibinfo{person}{Enrico Bertini}, {and} \bibinfo{person}{Andrea~G Parker}.}
  \bibinfo{year}{2021}\natexlab{}.
\newblock \showarticletitle{Mapping the Landscape of COVID-19 Crisis
  Visualizations}. In \bibinfo{booktitle}{\emph{Proceedings of the 2021 CHI
  Conference on Human Factors in Computing Systems}}. \bibinfo{pages}{1--23}.
\newblock


\end{thebibliography}
